\documentclass[5p]{elsarticle}
\usepackage{graphicx}
\usepackage{lineno}
\usepackage{kantlipsum,widetext}
\usepackage{hyperref}
\usepackage{color}
\usepackage[figuresleft]{rotating}
\usepackage{caption}
\usepackage{subcaption}
\usepackage{layout}
\usepackage{setspace}
\usepackage{dcolumn}
\usepackage{xtab}
\usepackage[T1]{fontenc}
\usepackage{textcomp}		
\usepackage[scaled=0.92]{helvet}
\usepackage{mathtools}
\usepackage{amsmath}
\usepackage{amssymb} 
\usepackage{multirow}
\usepackage{amsfonts}
\usepackage{tgtermes, tgheros}

\usepackage{array}
\usepackage{tabularx}
\newcolumntype{C}{>{\centering\arraybackslash}X}
\newcolumntype{L}{>{\raggedright\arraybackslash}X}
\newcolumntype{R}{>{\raggedleft\arraybackslash}X}
\usepackage[table]{xcolor}

\journal{International Journal of Production Economics}

\biboptions{authoryear}

\makeatletter
\def\ps@pprintTitle{%
\let\@oddhead\@empty
\let\@evenhead\@empty
\def\@oddfoot{}%
\let\@evenfoot\@oddfoot}
\makeatother
\begin{document}
\begin{frontmatter}

\title{
Structural propagation in a production network with \\ restoring substitution elasticities
}

\author[sn]{Satoshi Nakano}
\ead{nakano@jil.go.jp}

\author[kn]{Kazuhiko Nishimura\corref{cor1}
}
\ead{nishimura@n-fukushi.ac.jp}
\cortext[cor1]{Corresponding author}
\address[sn]{The Japan Institute for Labour Policy and Training, Tokyo 177-0044, Japan}
\address[kn]{Faculty of Economics, Nihon Fukushi University, Aichi 477-0031, Japan}


\begin{abstract}
We model an economy-wide production network by cascading binary compounding functions, based on the sequential processing nature of the production activities.  
As we observe a hierarchy among the intermediate processes spanning the empirical input--output transactions, we utilize a stylized sequence of processes for modeling the intra-sectoral production activities.
Under the productivity growth that we measure jointly with the state-restoring elasticity parameters for each sectoral activity, the network of production completely replicates the records of multi-sectoral general equilibrium prices and shares for all factor inputs observed in two temporally distant states. 
Thereupon, we study propagation of a small exogenous productivity shock onto the structure of production networks by way of hierarchical clustering. 
\end{abstract}
\begin{keyword}
Elasticity of Substitution \sep Productivity Growth \sep  Linked Input-Output Tables \sep Hierarchical Clustering
\end{keyword}
\end{frontmatter}

\section{Introduction \label{intro}}

Given the technological interdependencies among industrial activities, innovation (in terms of productivity shock) in one industry may well produce a propagative feedback effect on the performance of economy-wide production.
Previous literature pertaining to the study of innovation propagation has based its theory upon the non-substitution theorem \citep[e.g.,][]{ngr} that allows one to study under a fixed technological \text{structure} while restricting the 
analyses to changes in the net outputs \cite[e.g.,][]{pre2014, gvc}.
{Otherwise, \citet{acemoglu} assume Cobb-Douglas production (i.e., unit substitution elasticity) with which the structural transformation is restricted to the extent that the cost-share structures are preserved.} 
To study propagation in regard to potential technological substitutions, however, a potential range of alternative technologies must be known in advance. 
Nonetheless, empirical estimation of the substitution elasticities \cite[e.g.,][]{handbook} is elusive, and to this end, the dimension of the working variables has been significantly limited.

In contrast, this study is concerned with the economy-wide propagation of innovation that involves structural transformation with regard to the potential range of technologies among input variables of large (385) dimension. 
In so doing we model multiple-input production activity by serially nesting (i.e., cascading) binary-input production functions of different substitution elasticities. 
For each industry (or sector of an economy), the elasticity parameters of the production function are measured jointly with the productivity changes so that the economy-wide production system completely replicates monetary and physical inputs in all sectors for two temporally distant equilibrium states.
These elasticity parameters are hence called as restoring elasticities.

In the following, we give our rationale for  modeling production by nesting binary compounding functions.
Consider, say, the manufacture of a pair of jeans.
For this case, one needs a sufficient amount of denim fabric, a pair of scissors, a sewing machine, a ball of yarn, some electricity, and a tailor.  
We know that a pair of jeans will not fall into place all at once but rather it is made in a step-by-step fashion: using a pair of scissors and a sewing machine, the fabric is first cut into pieces, then they are sewn together using the yarn with some help from electric power.
Production generally involves a series of processes that combine the output of a previous process with another input of production, before handing the output over to the next process.

\begin{figure}[t!]
\includegraphics[width=.48\textwidth
]{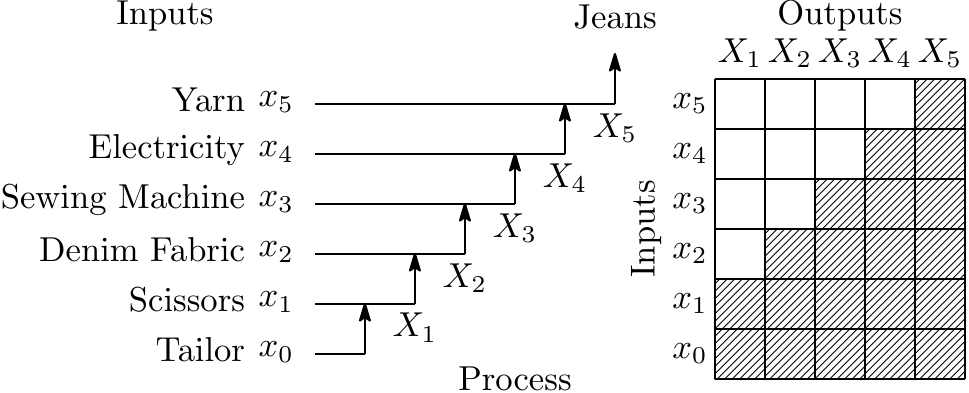}
 \caption{Serially nested configuration of a production system (left) and the corresponding incidence matrix (right) spanning direct and indirect inputs and intermediate outputs.\label{snc}}
 \end{figure}

A production activity can be configured as a tree diagram such as the one shown in Fig.~\ref{snc} (left).
In this example, the production system comprises a series of six inputs $(x_0$, $x_1$, $x_2$, $x_3$, $x_4$, $x_5)$, processed in a hierarchical manner, producing five intermediate outputs, $(X_1, X_2, X_3, X_4, X_5)$, by five processes that are nested serially.
Naturally, one may be concerned that the denim fabric, for example, is produced by another (satellite) system, and therefore the extended system is inclusive of a parallel nest.
Suppose, for simplicity, that denim fabric is produced by a serially nested process of two factor inputs, say, the indigo-dyed wrap and the plain weft threads, $(x_6, x_7)$.
Notice that we may always re-configure a production system into a sequence by decomposing the satellite process.
In this case, denim fabric is decomposed into a sequence of two inputs (wrap and weft) and a set of cut fabric (i.e., $X_2$) is, presumably, produced by the sequence (wrap, weft, tailor, scissors), in which case, the sequence of the inputs of the extended system becomes $(x_6, x_7, x_0, x_1, x_3, x_4, x_5)$.
\footnote{Of course, the weaving process of the threads requires a weaver.
Then, the same kind of input (labor) is put into process at different stages (weaving and tailoring).
By allowing indirect inputs we merge these inputs into the lower stage.}

A cascaded configuration of processes can be transcribed into a triangular incidence matrix, as shown in Fig.~\ref{snc}.
The shaded intersections represent the direct and indirect inputs and the distribution of outputs, while indirect feedbacks are ruled out for simplicity.\footnote{Indirect feedback may be, for example, the case where $X_5$ is fed back into $x_2$.  We rule this case out because there will be no way of producing the first pair of Jeans.}
A notable feature of this configuration is that every intermediate process constitutes a part in  the overall sequence of processes.
For instance, the intermediate process that produces output $X_3$ consists of four direct and indirect inputs, namely, $(x_0, x_1, x_2, x_3)$, that are processed in this sequence.
The overall sequence of processes is of the final output $X_5$, which is $(x_0, x_1, x_2, x_3, x_4, x_5)$, so the intermediate process $X_3$ obviously constitutes a part in  the overall sequence.
Thus, the sequence of every intermediate process is knowable by investigating the overall sequence of the triangular incidence matrix.

Our sector-level modeling of serially nested production activities requires the identification of the sequence of inputs (intra-sectoral processes) for all sectors constituting the economy, and for that matter, we utilize the economy-wide inter-sectoral transactions recorded in an input--output table.
We note that, if the incidence matrix transcribed from the input--output table is completely triangular, every sector-level sequence of processes becomes known from the sequence of intermediate processes stylized in the triangulated  incidence matrix.
We hereafter refer to this sequence as the universal processing sequence (UPS).
We will find that the input--output incidence matrix of the 2005 Japanese economy is not completely yet not too far from being triangular.
From an empirical perspective, we exploit the universal sequence of processes observed through the triangulation of the input--output incidence matrix.
Square matrix triangulation \cite[e.g., ][]{lop, lop2} refers to a generic technique for finding a simultaneous permutation of the rows and columns such that the sum of the entries above the main diagonal is maximized.

Given the hierarchical sequence of the inputs, we shall proceed to set up a production function that reflects the actual range of potential technologies.
In this study, we use the constant elasticity of substitution (CES) function \cite{acms} for modeling the binary compounding process.
CES has been applied exensively \citep[e.g.,][]{ijpeces}.
Nesting CES functions creates a Cascaded CES function whose nest elasticities are the main subject of estimation. 
As we show later on, the restoring nest elasticities will be resolved through calibration of the sectoral productivities, 
in order that the two temporally distant equilibrium states, regarding prices and shares of inputs spanning all sectoral productions, are replicated.
\footnote{Previous models are calibrated at one point \cite{rutherford}, while our model is subject to two-point calibration.}

The calibrated multi-sectoral Cascaded CES general equilibrium model is used to simulate the production networks transformation ex post of some external productivity shock, and account for its economy-wide influences in terms of welfare measured by the gains in the final demand.
Finally, we perform hierarchical cluster analysis upon the networks of production in different equilibria to study the potential structural propagation of the external productivity shock.
The clustering of production networks is studied in various ways \citep[e.g., ][]{pre2011, hu, sun} In this study, we base our cluster evaluation on the Leontief inverse which is particular case of Katz-Bonacich centrality \citep{carvalho}.
Upon performing hierarchical cluster analysis \citep[e.g., ][]{newman, jmn} we measure sectoral distances based on Pearson's correlation between sectoral multipliers of the changing production networks.


In section \ref{sec2}, we explain how the elasticity parameters and the productivity change of a Cascaded CES function are calibrated, given the universal processing sequence.
In section \ref{sec3}, we demonstrate that the equilibrium structures for both reference and current states are replicated, and how the production network 
is transformed by an exogenous productivity stimulus.
Section \ref{sec4} provides concluding remarks.
Note that superscripts are hereafter reserved for exponents, while subscripts are for indicating variety of inputs, nests, and industries, but not for partial derivatives.
The notation for key variables in different states 
is summarized in Table \ref{tab1}.
\begin{table}[t!]
\center
\small
\caption{Notations of variables and observations in different states (observations are highlighted).}
\label{tab1}
\begin{tabularx}{85mm}{lCCCC}
\hline
& Variable	& Reference 	& Current 	& Projected  	\\ \hline
Price	& $w$ & 1 & \cellcolor{gray!20}{$p$} & $\pi$ \\
Cost share &	$s$ & \cellcolor{gray!20}{$a$} & \cellcolor{gray!20}{$b$} & $m$ \\
Productivity & $t$ & 1 & $\theta$ & $\theta z$ \\
\hline
\end{tabularx}
\end{table}

\section{Model\label{sec2}}

\subsection{Cascaded CES Function \label{2a}}
Suppose that UPS of $n+1$ inputs is known.
Then, a cascaded (i.e., serially nested) production function of $n+1$ inputs for an industry (whose index $j$ is omitted) can be described as follows:
\begin{align*}
y =& {t} F \left( x_0, x_1, \cdots, x_{n} \right) = {t} F \left( \mathbf{x}, x_0 \right) \notag \\
=& {t} X_{n+1}\left( x_n, {X}_{n} \left( x_{n-1}, \cdots {X}_2\left( x_1, x_0 \right)  \cdots \right) \right)  
\end{align*}
Here, $y\geq0$ is the output, $x_i\geq0$ is the $i$th input, and ${t}>0$ is the productivity level.
There are $n$ nests, each consisting of one factor input and a compound from the lower level nest, except for the primary nest, which includes two non-compound inputs. 
Note that we allow $t$ to decrease (i.e., production activity may lose output performance instead of gaining), as may be observed in reality.
We define $X_1 = x_0$, for convenience.

The CES production function for the $i+1$th compound output processed at the $i$th nest is of the following form:
\begin{equation}
{X}_{i+1} \left( x_i, {X}_{i} \right) 
= \left( 
{\lambda}_{i}^\frac{1}{\sigma_{i}} 
x_{i}^\frac{\sigma_i-1}{\sigma_{i}} 
+ 
{\Lambda}_{i}
^\frac{1}{\sigma_{i}} 
{X}_{i}^\frac{\sigma_i-1}{\sigma_{i}} 
\right)^\frac{\sigma_{i}}{\sigma_i-1}
\label{vncespf}
\end{equation}
The nest production function (\ref{vncespf}) holds  for $i=1, \cdots,n$, since there are $n$ nests in a production activity.
Here, ${\lambda}_i = 1-\Lambda_i \in [0, 1]$ is the share parameter of the $i$th nest,
${X}_i$ is the compound output from the $i-1$th nest, and 
$\sigma_i$ is the elasticity of substitution between $i$th input $x_i$ and the compound input from the lower level nest ${X}_{i}$.
We assume that (\ref{vncespf}) is homogeneous of degree one. 

The following is the dual (or unit cost) function of the $i+1$th compound output given in (\ref{vncespf}):
\begin{align}
W_{i+1} \left( w_i, W_{i} \right) 
= \left( 
{\lambda}_{i} w_{i}^{1-\sigma_i} 
+ \Lambda_{i} 
W_{i}^{1-\sigma_i} 
\right)^\frac{1}{1-\sigma_i}
\label{vncesucf}
\end{align}
Here, $w_i$ is the price of the $i$th input, and $W_{i}$ is the price of the compound input from the lower level nest.
We may verify that (\ref{vncesucf}) is a dual function of (\ref{vncespf}) by the following exposition.
First, as we presume zero profit in the nest process, the following identity must hold: 
\begin{align}
W_{i+1} X_{i+1} = w_i x_i + W_{i} X_{i}
\label{iden}
\end{align}
Then, by virtue of (\ref{vncespf}) and (\ref{iden}), we have
\begin{align}
{\frac{\partial X_{i+1}}{\partial x_i}}/{\frac{\partial X_{i+1}}{\partial X_{i}}}
=\left( \frac{\lambda_{i}}{\Lambda_{i}}  \frac{X_{i}}{x_i} \right)^{1/\sigma_i}
=\frac{w_i}{W_{i}}
\label{mrts}
\end{align}
Alternatively, by virtue of (\ref{vncesucf}) and (\ref{iden}), we have
\begin{align}
{\frac{\partial W_{i+1}}{\partial w_i}}/{\frac{\partial W_{i+1}}{\partial W_{i}}}
= \frac{\lambda_{i}}{\Lambda_{i}} \left( \frac{W_{i}}{w_i} \right)^{\sigma_i}
=\frac{x_i}{X_{i}}
\label{duality}
\end{align}
Hence, it is safe to say that (\ref{vncesucf}) is the unit cost function of (\ref{vncespf}), since (\ref{mrts}) and (\ref{duality}) are equivalent.
A Cascaded CES unit cost function can then be created by nesting (\ref{vncesucf}) serially as follows:
\begin{align}
{c} &= t^{-1}H \left( w_0, w_1, \cdots, w_{n} \right) = t^{-1}H \left(\mathbf{w}, w_0 \right) 
\notag 
\\
&=  t^{-1}W_{n+1}\left( w_n, W_{n} \left( w_{n-1}, \cdots W_2\left( w_1, w_0 \right) \cdots \right)\right) \label{cn}
\end{align}
where $c$ indicates the unit cost of the sector concerned.
For convenience, we may define that $W_1 = w_0$.


\subsection{Restoring Elasticities \label{calib}}
A worked example of the calibration procedure that we present in this section for a two-stage Cascaded CES function is given in \ref{app-b}.
To begin with, let us apply Euler's homogeneous function theorem and the no-arbitrage (i.e., zero profit) condition to the unit cost function:
\begin{align*}
&c = \sum
\frac{\partial c}{\partial w_i} w_i,
&c = \sum
\frac{w_i x_i}{y} 
\end{align*}
By recursively taking the derivatives for  (\ref{vncesucf}), we arrive at the following identities for $i=0, 1, \cdots, n$:
\begin{align*}
&\frac{\partial W_{i+1}}{\partial w_i} = \lambda_i \left(\frac{W_{i+1}}{w_i}\right)^{\sigma_i},
&\frac{\partial W_{i+1}}{\partial W_i} = \Lambda_i \left(\frac{W_{i+1}}{W_i}\right)^{\sigma_i}
\end{align*}
Then, the cost share of the input of the $k$th nest, counting from the outermost of the $n$ nests, which we denote by $s_{n-k}$, can be derived as a function of the prices of the compound inputs, as follows:
\begin{align}
s_{n-k} 
&= \frac{x_{n-k}w_{n-k}}{yc}
=\frac{\partial c}{\partial w_{n-k}} \frac{w_{n-k}}{c} \notag \\
&=\frac{\partial W_{n+1}}{\partial W_n} \cdots \frac{\partial W_{n-k+2}}{\partial W_{n-k+1}}\frac{\partial W_{n-k+1}}{\partial w_{n-k}} \frac{w_{n-k}}{W_{n+1}} \notag \\
&= \lambda_{n-k} w_{n-k}^{1-\sigma_{n-k}} W_{n+1}^{\sigma_n -1}\prod_{l=0}^{k-1} \Lambda_{n-l} W_{n-l}^{\sigma_{n-l-1} -\sigma_{n-l}} \label{snk}
\end{align}

Our task here is to solve for $\sigma_i$ for all $i$ and the change in $t$, using the shares and prices observed in two different states, namely, the current and the reference. 
First, we standardize all prices by those of the inputs (and outputs) of the reference state and set them at unity, i.e., $(w_0, \mathbf{w})=(1, \mathbf{1})$, while denoting the reference-standardized prices of the current state by $\mathbf{p}=(p_0, p_1, \cdots, p_n)$.   
The productivity level for the reference state must also be standardized and set to unity, i.e., $t=1$, since we know from (\ref{vncesucf}) that $W_{i+1}=w_i=1$ for all $i$.
We also let $p_0$ be given outside of the system, because the primary input (i.e., num{\'e}raire good or labor) is not produced industrially.

We denote the observed cost share of the $i$th input of the reference state by $a_i$ and that of the current state by $b_i$.
For later convenience, we introduce $\chi_i = \left\{ a_i, b_i, p_i \right\}$, a set of observables in the two states for the $i$th input.
Further, we note that  $\chi_{n+1}=\left\{p\right\}$ and $\chi_{0}=\left\{p_0\right\}$ for convenience.
Applying reference state values into (\ref{snk}) yields the following:
\begin{align}
a_{n-k} =\lambda_{n-k} \prod_{l=0}^{k-1} \Lambda_{n-l}
\label{ank}
\end{align}
The following modification may not be so obvious; nonetheless, we may see the equivalence by applying all possible $k$ recursively into (\ref{ank}).
By taking $\Lambda_i=1-\lambda_i$ into account, 
Further (\ref{ank}) can be modified to calibrate the share parameters:
\begin{align}
\lambda_{n-k} = \frac{a_{n-k}}{1-\sum_{l=0}^{k-1} a_{n-l}}
\label{lamnk}
\end{align}
Applying current state values to (\ref{snk}) 
we have:
\begin{align*}
b_{n-k} = a_{n-k} p_{n-k}^{1-\sigma_{n-k}} W_{n+1}^{\sigma_n -1} \prod_{l=0}^{k-1}W_{n-l}^{-\sigma_{n-l} + \sigma_{n-l-1}}
\end{align*}
By rearranging terms, we obtain the following:
\begin{align*}
\sigma_{n-k} = \frac{\ln \frac{b_{n-k}}{a_{n-k}} + \sum_{l=0}^{k-1} \sigma_{n-l} \ln \frac{W_{n-l}}{W_{n-l+1}} + \ln \frac{W_{n+1}}{p_{n-k}}  }{\ln \frac{W_{n-k+1}}{p_{n-k}}}
\end{align*}
We write the above in terms of its entries (unknowns and knowns separated by a semi-colon) for convenience:
\begin{align}
\sigma_{n-k} \left( W_{n-k+1}, \cdots, W_{n+1}, \sigma_{n-k+1}, \cdots, \sigma_{n}; \chi_{n-k} \right)
\label{sigmank}
\end{align}
Also, (\ref{vncesucf}) is 
evaluated at the current state:
\begin{align*}
W_{n-k+1} = \left( \frac{W_{n-k+2}^{1-\sigma_{n-k+1}}-b_{n-k+1}p_{n-k+1}^{1-\sigma_{n-k+1}}}{1-b_{n-k+1}} \right)^{\frac{1}{\sigma_{n-k+1}}}
\end{align*}
Again, we write the above in terms of its entries:
\begin{align}
W_{n-k+1} \left( W_{n-k+2}, \sigma_{n-k+1}; \chi_{n-k+1} \right)
\label{wnk1}
\end{align}

Now, let us work on  (\ref{sigmank}) and (\ref{wnk1}) step by step from the outer layer of the nests, i.e., $k=0$.
For this particular layer, we may resolve the unknowns, given $t$, viz.,
\begin{align*}
&\sigma_n\left(t; \chi_{n+1}, \chi_n \right) = \frac{\ln \frac{b_n W_{n+1}}{a_n p_n}}{\ln \frac{W_{n+1}}{p_n}},
&W_{n+1}\left(t; \chi_{n+1}\right) = tp
\end{align*}
Note that the second equation is restating (\ref{cn}) at the current state $c=p$.
Using the above terms the compound price $W_n$ is evaluated as follows:
\begin{align*}
W_n 
&= \left( \frac{(tp)^{\frac{\ln a_n - \ln b_n}{\ln tp - \ln p_n}} - b_n p_n^{\frac{\ln a_n - \ln b_n}{\ln tp - \ln p_n}}}{1-b_n} \right)^{\frac{\ln tp - \ln p_n}{\ln a_n - \ln b_n}} 
\\
&=W_n\left( t; \chi_{n+1}, \chi_{n} \right)
\end{align*}
We may work on a few layers below:
\begin{flalign*}
&W_n\left( W_{n+1}, \sigma_n; \chi_n \right)  
\Rightarrow
W_{n}\left(t; \chi_{n+1}, \chi_n \right)
\\
&\sigma_{n-1}\left( W_n, W_{n+1}, \sigma_n; \chi_{n-1} \right)
\Rightarrow
\sigma_{n-1}\left( t; \chi_{n+1}, \chi_{n},  \chi_{n-1} \right)
\\
&W_{n-1}\left( W_{n}, \sigma_{n-1}; \chi_{n-1} \right) 
\Rightarrow
W_{n-1}\left(t; \chi_{n+1}, \chi_n, \chi_{n-1}\right)
\\
&\sigma_{n-2}\left( W_{n-1}, W_{n}, W_{n+1}, \sigma_{n-1}, \sigma_{n}; \chi_{n-2} \right) \\
&~~~~~~~~~~~~~~~~~~~~~~~~~~~~~\Rightarrow
\sigma_{n-2}\left( t; \chi_{n+1}, \chi_{n}, \chi_{n-1},  \chi_{n-2} \right)  
\end{flalign*}
We repeat this procedure and obtain the following series:
\begin{align*}
&W_{n-k+1}\left( t; \chi_{n+1}, \chi_{n}, \cdots, \chi_{n-k+1} \right) \\
&\sigma_{n-k}\left( t; \chi_{n+1}, \chi_{n}, \cdots, \chi_{n-k} \right)
\end{align*}
Thus, $t$ can be calibrated by way of the condition of the final stage $k=n$ at the current state, i.e.,
\begin{align}
W_{1}\left( t; \chi_{n+1}, \chi_n, \cdots, \chi_1 \right) = p_0
\label{w1p0}
\end{align}
and the restoring elasticity parameters $\sigma_{n-k}$ can all be resolved for $k=0, 1, \cdots, n-1$, using the solution $t$ of (\ref{w1p0}), which we hereafter denote by $\theta$.

\section{Empirical Analysis \label{sec3}}
\subsection{Universal Processing Sequence \label{oon}}
The degree to which a macroscopic production structure agrees with the hierarchical order of processing sequences is the \textit{linearity}.
In a perfectly linear structure, the processing sequences will only cascade from upstream to downstream;
if this is the case, then we may arrange the rows and columns of the input--output matrix according to the hierarchical order to obtain a triangular matrix, and hence, the universal processing sequence. 

More specifically, for an $n$ sector output system with $n-1$ intermediate inputs (excluding self-input), the furthest upstream (headwaters) sector has no intermediate input and $n-1$ output destinations (i.e., zero column and $n-1$ row entries)  whereas the furthest downstream sector has $n-1$ inputs with no intermediate output destination (i.e., $n-1$ column and zero row entries).
Let us denote a reordering of $n$ sectors whose initial order is $\left(1, 2, \cdots, n\right)$ by a permutation mapping $\phi: \left(\phi(1),\phi(2),\cdots,\phi(n)\right)$.
Further, let $k(\phi)= \left\{ l ~|~ k=\phi(l)   \right\}$ designate the inverse, so that a $\phi$-permuted version of a matrix $\mathbf{U}=\left\{ u_{ij} \right\}$ can be specified as follows:
\begin{align*}
\mathbf{U} \left( \phi \right) 
= \left\{ u(\phi)_{ij} \right\} 
= \left\{ u_{i(\phi)j(\phi)} \right\}
\end{align*}
For later discussion, let us work on  a discretized square input--output matrix $\mathbf{U}$ (i.e., input--output incidence matrix) whose elements are binary, specifically, $u_{ij}=1$ if transaction $x_{ij} \neq 0$, and $u_{ij}=0$ if $x_{ij} = 0$.
The linearity $\ell$ of a matrix $\mathbf{U}$ with permutation $\phi$ is defined as follows:
\begin{align*}
\ell
= \frac{\sum_{i<j} u\left({\phi}\right)_{ij}}{\sum_{i \neq j} u\left({\phi}\right)_{ij}} 
=\frac{h\left( \mathbf{U} \left(\phi \right) \right)}{K}
\end{align*}
Note that the denominator $K$ is the sum of the off-diagonal entries of the matrix, which is constant.
The numerator $h\left( \mathbf{U} \left(\phi \right) \right)$, on the other hand, depends on the permutation.

The triangulation problem of matrix $\mathbf{U}$ is to find a permutation mapping $\phi$ that maximizes the linearity, i.e., 
\begin{align*}
\max_{\phi \in \Phi} \, \ell = \frac{h\left( \mathbf{U} \left(\phi \right) \right)}{{K}}
\end{align*}
where $\Phi$ is the set of all possible permutations.
This problem is known as the linear ordering problem. 
The number of all possible permutations is as large as $n!$ for an $n \times n$ matrix, and the problem is known to be NP-hard.
Thus, none of the exact methods (typically via discrete optimization) will work when one attempts to handle a matrix of 385 sectors. 
Hence we take a heuristic approach.

One possible approach would be to use the ratio of the input incidents total (column sum of $\mathbf{U}$) to the output incidents total (row sum of $\mathbf{U}$), which we denote by $z_1$ and define below, and to arrange the permutation in the {ascending} order of this ratio.
\begin{align*}
z_1(k) = \frac{\sum_{i=1}^n u_{ik}}{\sum_{j=1}^n u_{kj}}
\end{align*}
This heuristic is a simple emulation of the method proposed by \citet{cw}, except that the original study uses the input--output coefficient matrix instead of an incidence matrix.
Note that in this case the set of possible permutations contains only one element, which we denote by $\phi_1$; that is, $\Phi=\left\{ \phi_{1} \right\}$. 
In other words, permutation $\phi_{1}: \left(\phi_1(1), \phi_1(2), \cdots, \phi_1(n)\right)$
is the ascending order of the above-mentioned ratios $z_1: \left( z_1(1), z_1(2), \cdots, z_1(n) \right)$, or $z_1(1(\phi_1)) \leq z_1(2(\phi_1)) \leq \cdots \leq z_1(n(\phi_1))$.

In this study, we slightly further generalize this  CW heuristic.
In particular, we take the \textit{weighted} ratio of the input incidents total to the output incidents total, as described below, in order to expand the number of possible permutations ${\Phi}$.
\begin{align*}
z_{\gamma}(k) = \frac{\left(\sum_{i=1}^n u_{ik}\right)^{\gamma}}{\sum_{j=1}^n u_{kj}}
\end{align*}
This ratio is inclusive of CW heuristic at $\gamma =1$.
We thus evaluate the linearity of $\mathbf{U}$ with permutation $\phi_{\gamma}: \left(\phi_{\gamma}(1), \phi_{\gamma}(2), \cdots, \phi_{\gamma}(n)\right)$
with respect to the ascending order of the ratios $z_{\gamma}: \left( z_{\gamma}(1), z_{\gamma}(2), \cdots, z_{\gamma}(n) \right)$, such that $z_{\gamma}(1(\phi_{\gamma})) \leq z_{\gamma}(2(\phi_{\gamma})) \leq \cdots \geq z_{\gamma}(n(\phi_{\gamma}))$ for all $\gamma \in \Gamma$.

Note that $\gamma$ must be nonnegative, according to the purpose of triangulating matrix $\mathbf{U}$. Obviously, $\gamma \in [0,1)$ if we are  to put more weight on the outputs, and $\gamma \in (1,\infty)$ if we are  to put more on the inputs.
Our objective is hence to search for the $\gamma$ that maximizes the linearity, 
that is, to find a $\gamma^*$ such that 
\begin{align}
\max_{\gamma \in \Gamma} \, \ell = \frac{h\left( \mathbf{U} \left(\phi_{\gamma} \right) \right)}{K}
\label{maxlgamma}
\end{align}
We shall hereinafter refer to the order of sectors given by $\phi_{\gamma*}$ as the \textit{stream order}.

\subsection{Data and Measurement}
\begin{figure*}[t!]
\includegraphics[width=.48\textwidth]{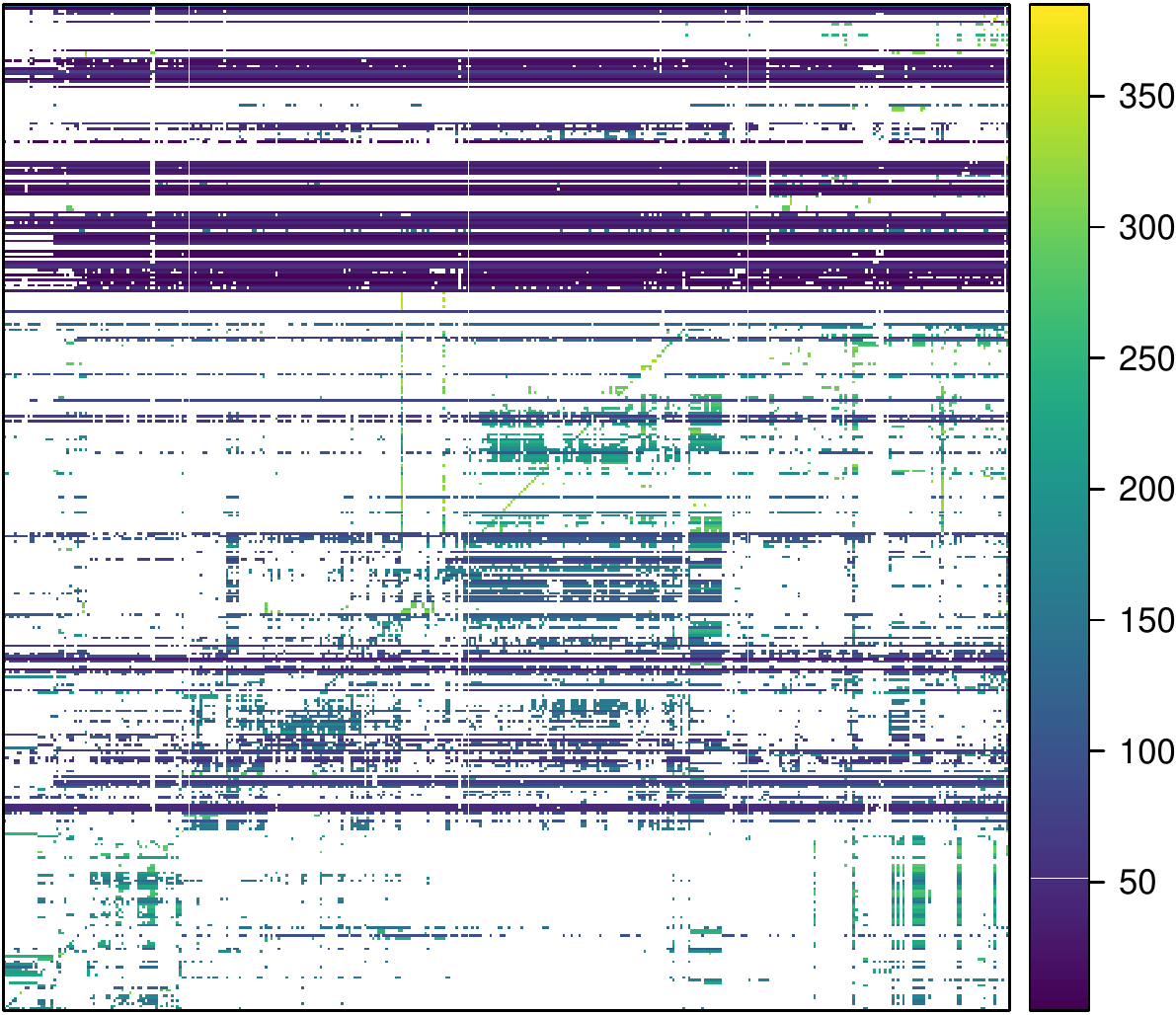}%
\hspace{3mm}
\includegraphics[width=.48\textwidth]{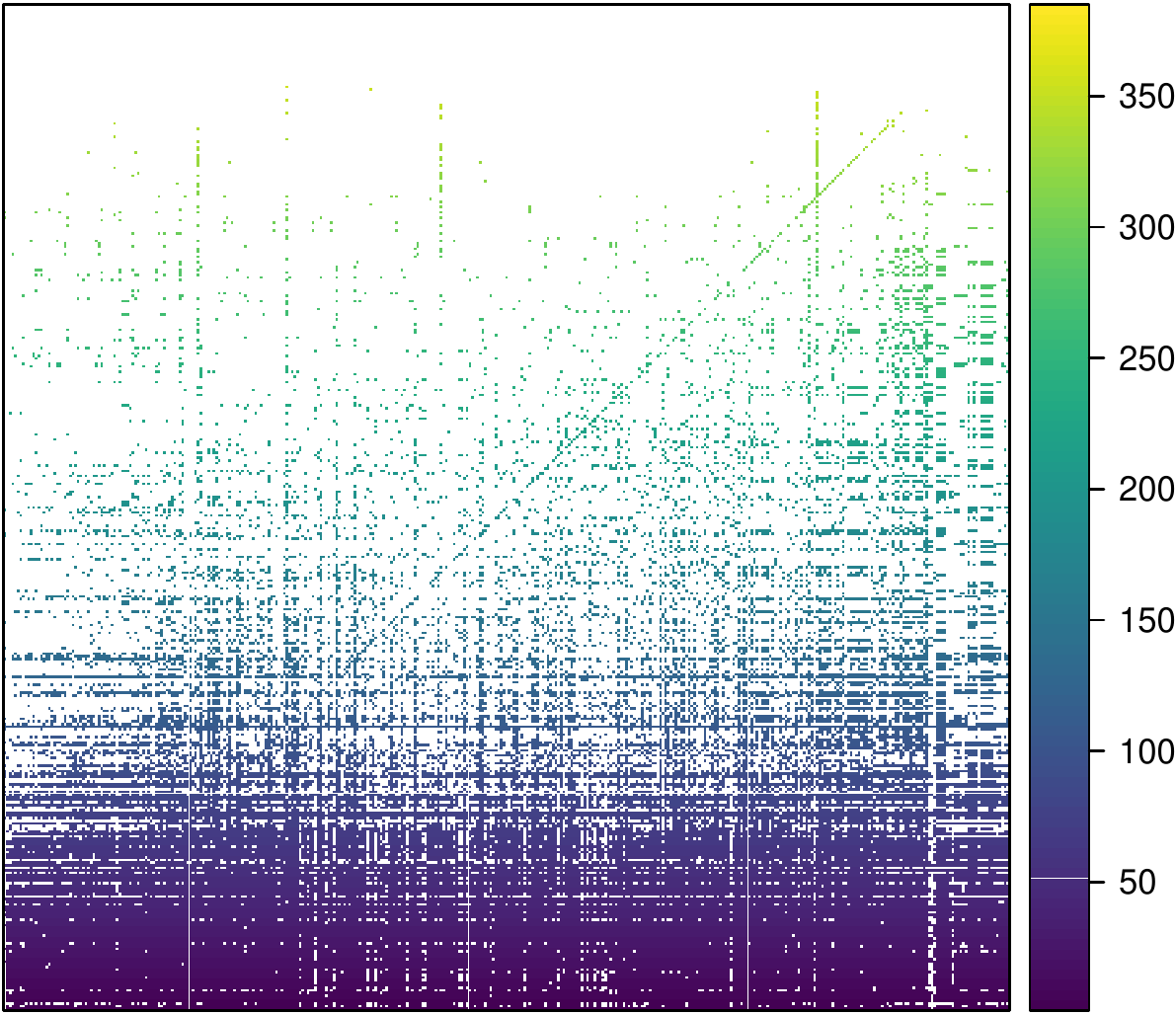}%
\caption{Input--output incidence matrix $\{ u_{ij} \}$ for Japan (based on 2005--2011 averaged transactions) of the original order (left) and the stream order (right).
The incidences are colored according to the position in the stream order of inputs in both cases.
The stream order is obtained by the linearity maximization via the generalized CW heuristic  as illustrated in Fig.~\ref{fig_linear}
\label{fig_triangle}}
\end{figure*}
\begin{figure}[t!]
\includegraphics[width=.49\textwidth]{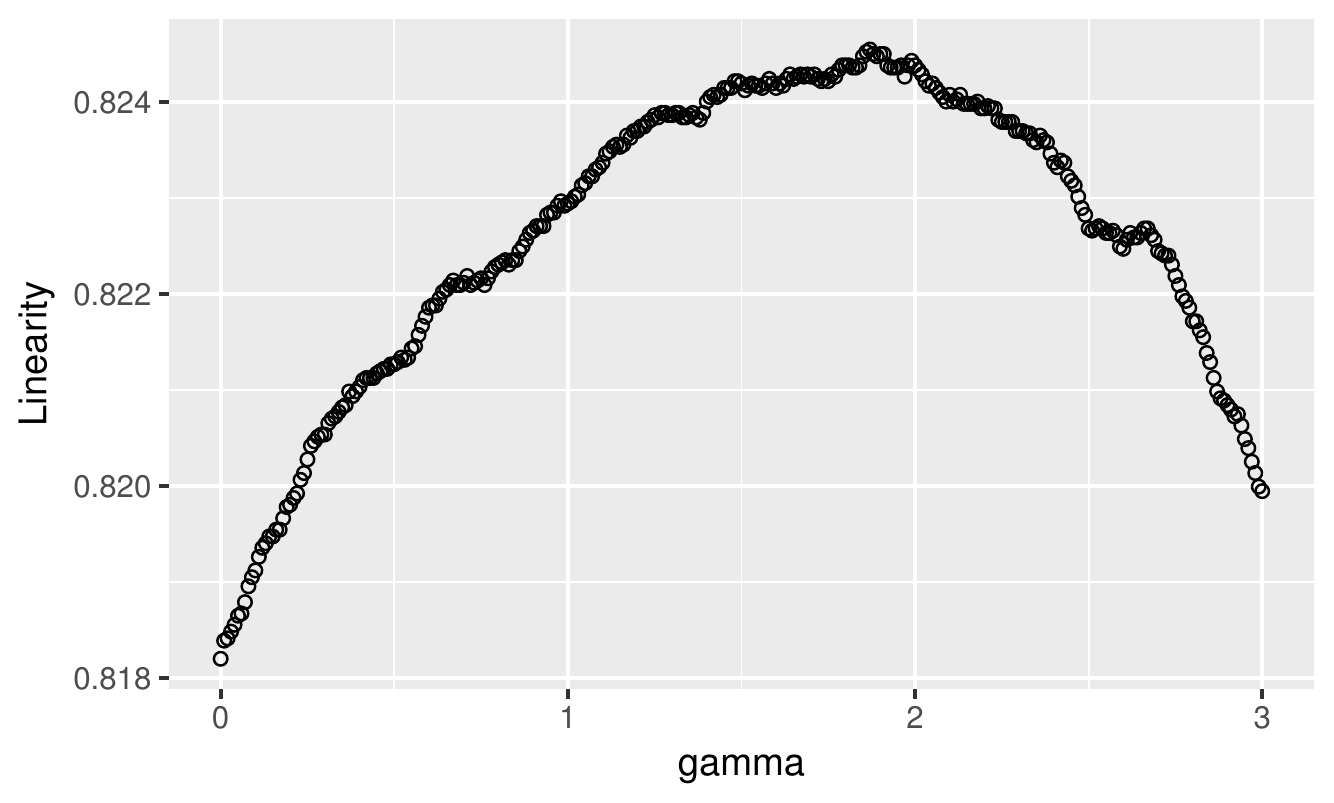}%
\caption{
Linearity $\ell$ of the input--output incidence matrix for Japan (based on 2005--2011 averaged transactions) is maximized at $\gamma^* = 1.87$ by the  generalized CW heuristics. \label{fig_linear}}
\end{figure}

We base our empirical study upon Japan's 2000--2005--2011 linked input--output tables \cite{miac}.
This set of tables includes the yearly commodity transactions of 385 factors (inputs) among the 385 industrial sectors (outputs) in Japan for three different periods.
The transactions are recorded in nominal monetary values; thus, in order to enable comparison in real (or physical) quantities between different periods, nominal values are converted into real values by a set of price indexes called deflators, which are  also included in a set of linked input--output tables.
We thus have three input--output transaction matrices which can be converted into a series of input shares and prices spanning the three periods of observation.

When we began to study this data set, we found a considerable number of inconsistent transactions.
Namely, 0.15\% of the incidents 
that existed in 2000 had disappeared by 2005, and 0.46\% of the incidents that existed in 2005 had disappeared by 2011.
Likewise, 0.54\% of the incidents that existed in 2005 did not exist in 2000, and 0.21\% of the incidents that existed in 2011 did not exist in 2005.
Such inconsistencies in transactions interfere with our calibration of the parameters; thus, we decided to use the two intermediate values of the three observations.
In other words, we merged two temporally neighboring transactions and created 2000--2005 and 2005--2011 average transaction tables.
We still, however, observed inconsistencies between the two incidents for, for example, the case where $\left( x({2000}), x({2005}), x({2011}) \right) = \left( 0,0,x \right)$; merging the neighboring transactions, i.e., $\left( x({2000, 2005}), x({2005, 2011}) \right) = \left( 0,x \right)$, cannot circumvent this inconsistency.
In such cases, to avoid inconsistencies we modified the transactions such that $\left( x({2000, 2005}), x({2005, 2011}) \right) =(0.25x, 0.75x)$.
Such modification was needed for 0.2\% of the 2005--2011 non-zero transactions.
As for 
prices, we used the mean values of the two temporally neighboring deflators for each commodity.

Figure~\ref{fig_linear} shows the result of performing the search for $\gamma$ in $\Gamma=[0, 3]$ using the incidence matrix based on 2005--2011 average input-output transactions.
The solution of (\ref{maxlgamma}) is $\ell=0.8245$ at $\gamma=1.87$.
Note that $\ell=0.8229$ for the CW heuristic, which occurs at $\gamma=1$. Figure \ref{fig_triangle} shows how the input--output incidence matrix is actually triangulated with our heuristic under $\gamma=1.87$.
We find that sectors such as Office supplies, Water supply, Gas supply, Electricity, Construction services, and Information services, are among the upstream (at the top of the stream order), whereas Medical service, School education, Public baths, Barber shops, and Movie theaters, are among the downstream (at the bottom of the stream order).

\begin{figure}[t!]
\includegraphics[width=.47\textwidth]{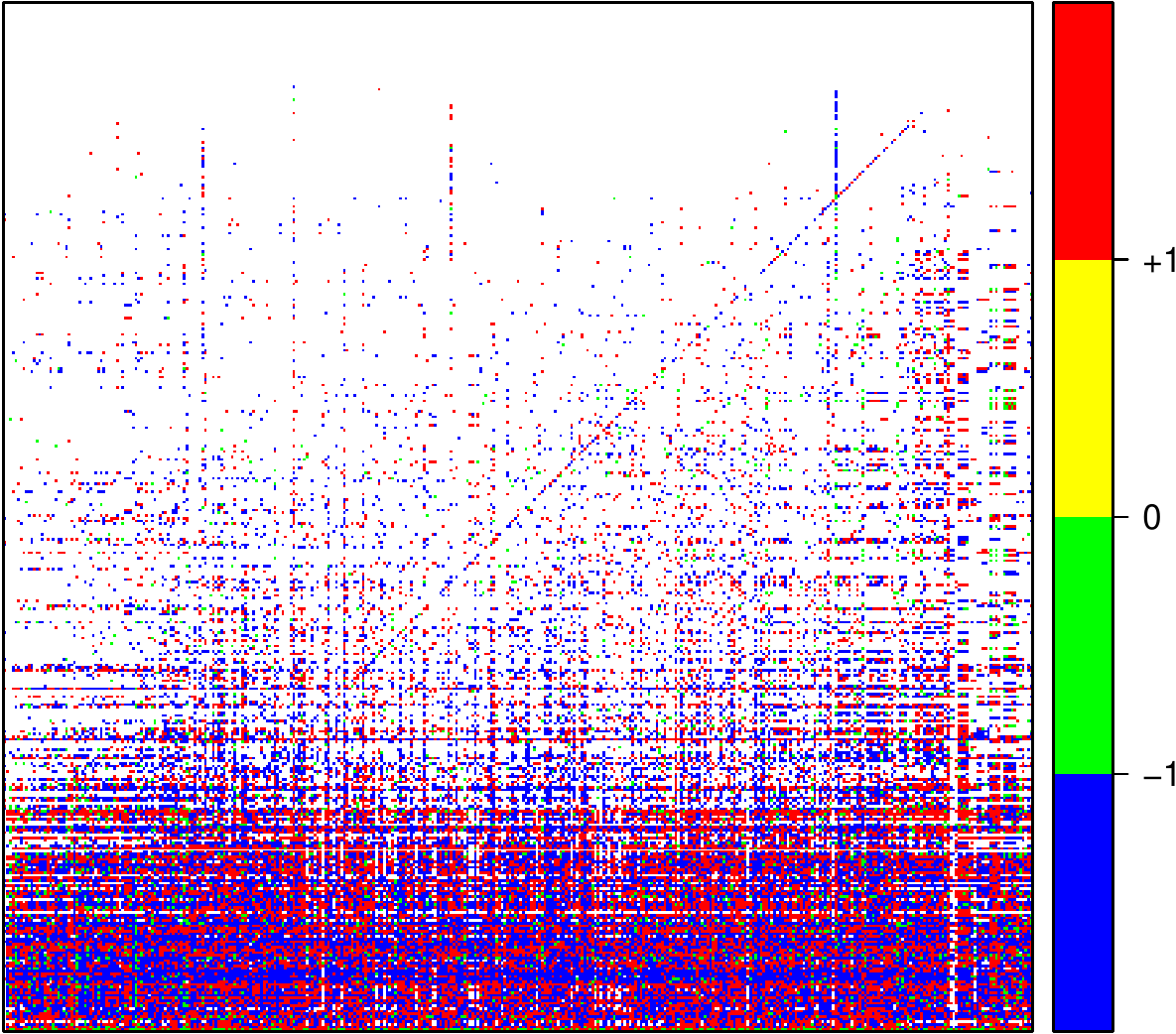}%
\caption{Elasticity parameters $\sigma_{ij}$ of Cascaded CES functions, displayed in stream order. \label{sigma}}
\end{figure}

\begin{figure}[t!]
\includegraphics[width=.49\textwidth]{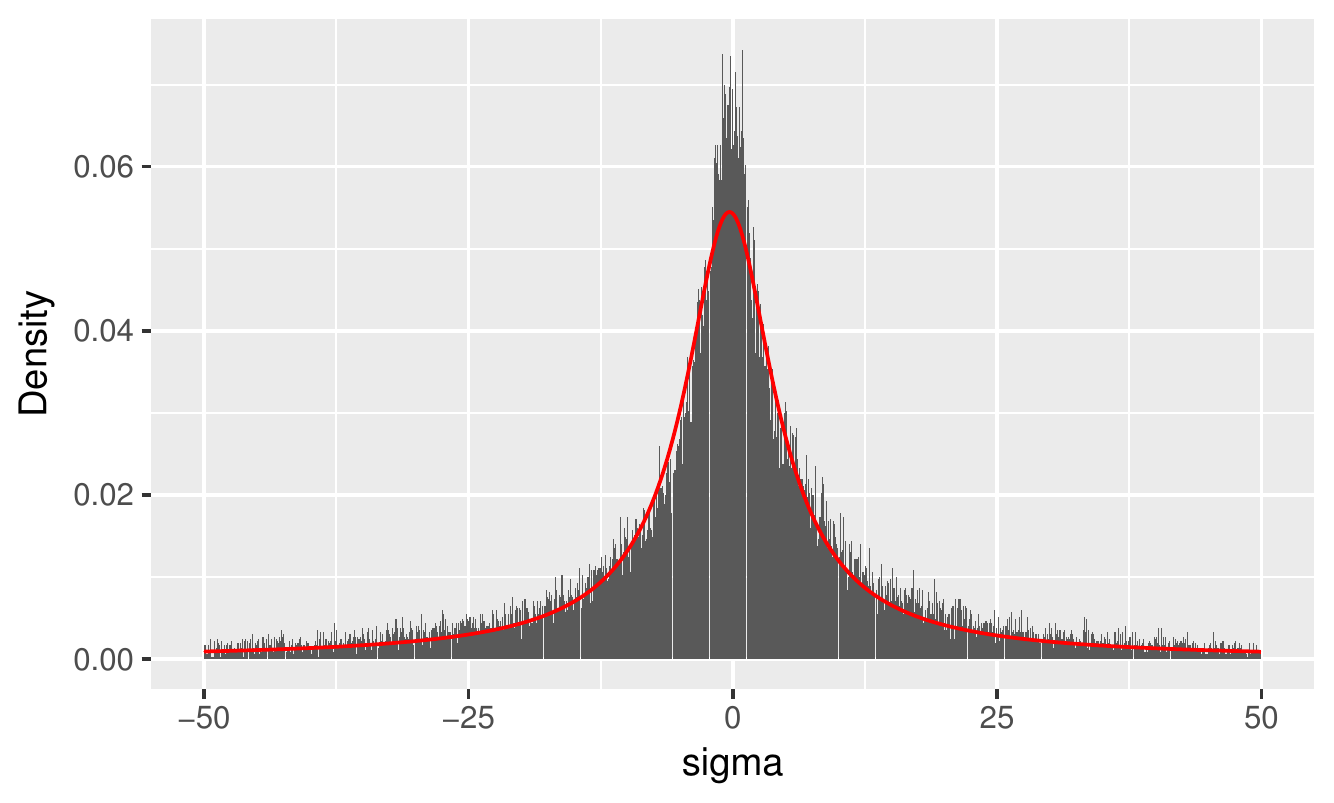}%
\caption{Distribution of Cascaded CES elasticity parameters $\sigma_{ij}$ (in the indicated truncated range).  
The histogram is fitted by the q-Gaussian distribution with $q=2.134$. 
\label{sigma_dist}}
\end{figure}

\begin{figure*}[t!]
\includegraphics[width=0.98\textwidth]{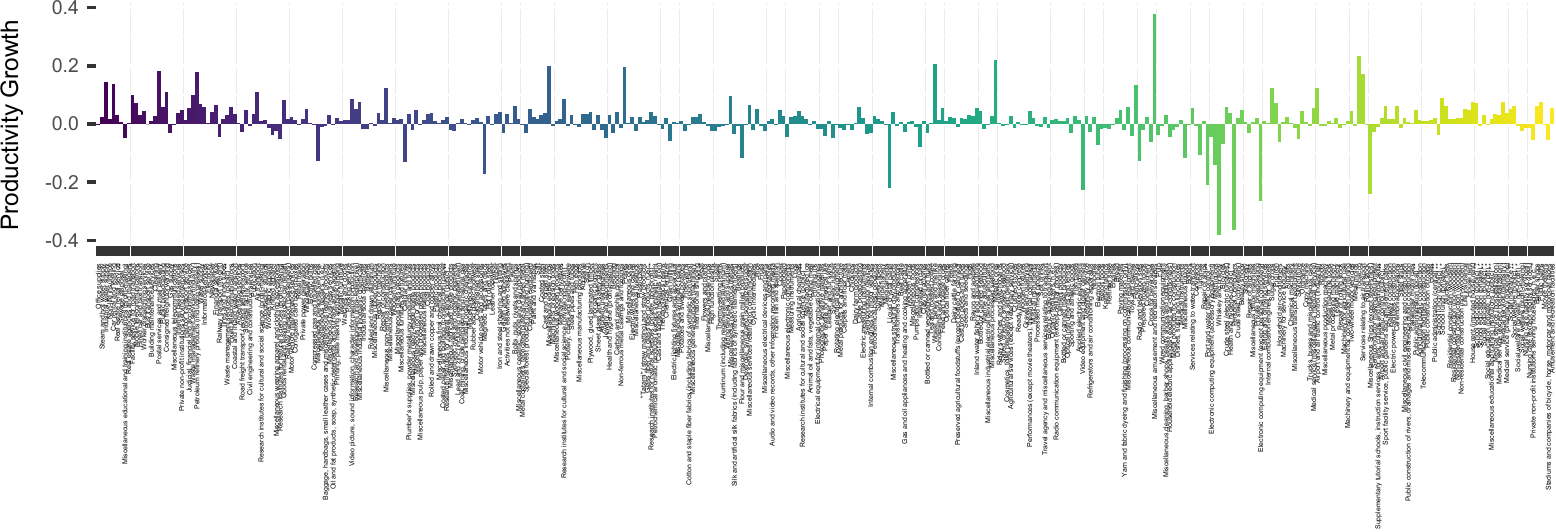}
\caption{Sectoral productivity growths measured under Cascaded CES functions, displayed in stream order. 
Sectors with positive productivity growths include Miscellaneous amusement and recreation services ($0.376$), Tobacco ($0.233$), and Metallic ores (0.219).
Sectors with negative productivity growths include Whiskey and brandy ($-0.383$), Personal Computers ($-0.364$), and Electronic computing equipment (except personal computers) ($-0.266$). 
These numbers are almost identical to the log of T{\"o}rnqvist indices obtainable via (\ref{tl}). 
 \label{tfp}}
\end{figure*}

Hereafter, we shall use the 2000--2005 average transactions as the \textit{reference} and the 2005--2011 average transactions as the \textit{current}.
Figure \ref{tfp} shows the results of a productivity calibration by solving (\ref{w1p0}) for all industries, using the reference and current input shares and deflators obtainable from the 2000--2005 average and 2005--2011 average transactions.
The solution of (\ref{w1p0}), i.e., $\theta_j$, is converted into a growth term with respect to the reference state, which we call the productivity growth $\ln \theta_j$.
Notice that these productivity growths were almost identical to the following widely used productivity measure (i.e., total factor productivity growth, TFPg) based on T{\"o}rnqvist index. 
\begin{align}
\text{TFPg} = -\ln p + \sum_{i=0}^n \left( \frac{a_{ij} + b_{ij}}{2} \right) \ln p_i
\label{tl}
\end{align}
\citet{diewert} showed that (\ref{tl}) can exactly capture the productivity growth of the underlying translog function.
In Fig.~\ref{sigma}, we show the elasticity parameters for individual nests, resolved by way of  (\ref{sigmank}), (\ref{wnk1}), and (\ref{w1p0}).
The elasticity parameters are represented by colors: yellow is used for $0 \leq \sigma_{ij} <1$, red for $1 \leq \sigma_{ij} < +\infty$, blue for $-\infty < \sigma_{ij} \leq -1$, and green for $-1 < \sigma_{ij} <0$, as indicated in the right bar. 
The distribution of the elasticity parameters is shown in Fig.~\ref{sigma_dist}.
Note that the histogram is fitted by the q-Gaussian distribution \cite{qg}.
The main parameter was estimated to be $q=2.134$ with standard error $ =0.0064$ using the method established by \citet{qgr}, while other parameters were fitted at $q$-mean $=-0.31$ and $q$-variance $=5.52$ by minimizing the residual sum of squares.



\subsection{Multisectoral General Equilibrium}
Let us now verify that the productivity growths we measured correctly replicate the two states, reference and current, in the light of multisectoral general equilibrium.
We begin by writing (\ref{cn}) in vector form:
\begin{align}
\left(c_1, \cdots, c_n\right)
&=\left( t^{-1} H_1\left( \mathbf{w}, w_0 \right), \cdots, t^{-1} H_n\left( \mathbf{w}, w_0 \right)\right) \notag \\
\mathbf{c} 
&= \mathbf{H}\left( \mathbf{w}, w_0 \right) \left< \mathbf{t} \right>^{-1}
\label{ch}
\end{align}
where $\mathbf{H}$ denotes a row vector of Cascaded CES unit cost functions and angle brackets indicate a diagonalized vector.
The gradient of $\mathbf{c}$ is
\begin{align*}
\nabla \mathbf{{c}} 
&=
\begin{bmatrix}
\frac{\partial {t}_1^{-1}{H}_1\left( \mathbf{w}, w_0\right)}{\partial w_0} & \frac{\partial {t}_2^{-1}{H}_2\left( \mathbf{w}, w_0\right)}{\partial w_0} & \cdots & \frac{\partial {t}_n^{-1}{H}_n\left( \mathbf{w}, w_0\right)}{\partial w_0}  \\
\frac{\partial {t}_1^{-1}{H}_1\left( \mathbf{w}, w_0\right)}{\partial w_1} & \frac{\partial {t}_2^{-1}{H}_2\left( \mathbf{w}, w_0\right)}{\partial w_1} & \cdots & \frac{\partial {t}_n^{-1}{H}_n\left( \mathbf{w}, w_0\right)}{\partial w_1}  \\
\vdots & \vdots & \ddots & \vdots  \\
\frac{\partial {t}_1^{-1}{H}_1\left( \mathbf{w}, w_0\right)}{\partial w_n} & \frac{\partial {t}_2^{-1}{H}_2\left( \mathbf{w}, w_0\right)}{\partial w_n} & \cdots & \frac{\partial {t}_n^{-1}{H}_n\left( \mathbf{w}, w_0\right)}{\partial w_n}  
\end{bmatrix}
\\
&=
\begin{bmatrix}
\nabla_{0} \mathbf{H}\left( \mathbf{w}, w_0\right)  \\
\nabla \mathbf{H}\left( \mathbf{w}, w_0\right)
\end{bmatrix}
\left< \mathbf{t} \right>^{-1}
\end{align*}
where $\nabla_0\mathbf{H}$ is a row vector, and $\nabla\mathbf{H}$ is an $n\times n$ matrix.

According to Shephard's lemma, current state input shares $\left\{ b_{ij} \right\}=\left[ \mathbf{b}_0, \mathbf{B} \right]^\prime$ can be described in terms of the current state (i.e., $\mathbf{t}=\boldsymbol{\theta}$, $\mathbf{w}=\mathbf{p}$, $w_0 = p_0$) by the gradient of the unit cost function, as follows:
\begin{align*}
\mathbf{b}_0 &= p_0 \nabla_0 \mathbf{H}\left( \mathbf{p}, p_0 \right) \left< \boldsymbol{\theta}\right>^{-1} \left< \mathbf{p} \right>^{-1} \\
\mathbf{B}&= \left< \mathbf{p}\right> \nabla \mathbf{H}\left( \mathbf{p}, p_0 \right) \left<\boldsymbol{\theta}\right>^{-1}\left< \mathbf{p} \right>^{-1} 
\end{align*}
Likewise, for the reference state (i.e., $\mathbf{t}=\boldsymbol{1}$, $\mathbf{w}=\mathbf{1}$, $w_0 = 1$), the following must hold for the reference input shares $\left\{ a_{ij} \right\}=\left[ \mathbf{a}_0, \mathbf{A} \right]^\prime$:
\begin{align*}
\mathbf{a}_0 &= 1 \nabla_0 \mathbf{H}\left( \mathbf{1}, 1 \right) \mathbf{I}^{-1} \mathbf{I}^{-1} = \nabla_0 \mathbf{H}\left( \mathbf{1}, 1 \right) \\
\mathbf{A}&= \mathbf{I} \nabla \mathbf{H}\left( \mathbf{1}, 1 \right) 
\mathbf{I}^{-1} \mathbf{I}^{-1} =\nabla \mathbf{H}\left( \mathbf{1}, 1 \right)
\end{align*}
Note that all the parameters of $\mathbf{H}$, i.e., $\sigma_{ij}$, have been resolved previously, to satisfy the above four equations for the two states, through the calibration of $\boldsymbol{\theta}$.
Thus, we verify that $\mathbf{t}=\mathbf{1}$ and $\mathbf{t}=\boldsymbol{\theta}$ can actually replicate the two equilibrium prices $\mathbf{c}=\mathbf{w}=\mathbf{1}$ and $\mathbf{c}=\mathbf{w}=\mathbf{p}$, with $p_0$ given.

Because $\mathbf{H}$ is homogeneous of order one in $(\mathbf{w}, w_0 )$, Euler's homogeneous function theorem is applicable for (\ref{ch}). 
At the reference state, the theorem implies that
\begin{align*}
\mathbf{H}\left( \mathbf{1}, 1 \right) \left<\boldsymbol{1}\right>^{-1}
&=\left[ 1 \nabla_0 \mathbf{H}\left( \mathbf{1}, 1 \right)  
+\mathbf{1} \nabla \mathbf{H}\left( \mathbf{1}, 1 \right) \right] \mathbf{I}^{-1} 
\\
&=\mathbf{a}_0 \mathbf{I}
+ 
\mathbf{1}\mathbf{I}^{-1}
\mathbf{A}\mathbf{I}
\\
&= \left[ \mathbf{a}_0 + \mathbf{1} \mathbf{A} \right] 
\mathbf{I}
=\mathbf{1}\mathbf{I}
=\mathbf{1}
\end{align*}
Note that the last part of the equality is due to the nature of shares: $\sum_{i=0}^n s_{ij} =1$. 
Likewise, similar result is obtainable for the current state, as follows:
\begin{align*}
\mathbf{H}\left( \mathbf{p}, p_0 \right) \left<\boldsymbol{\theta}\right>^{-1}
&=\left[ p_0 \nabla_0 \mathbf{H}\left( \mathbf{p}, p_0 \right) 
+\mathbf{p} \nabla \mathbf{H}\left( \mathbf{p}, p_0 \right) \right]
\left<\boldsymbol{\theta}\right>^{-1} 
\\
&=\mathbf{b}_0 \left< \mathbf{p}\right>
+ \mathbf{p} \left< \mathbf{p} \right>^{-1}\mathbf{B}\left< \mathbf{p} \right>
\\
&= \left[ \mathbf{b}_0 + \mathbf{1} \mathbf{B} \right] \left<\mathbf{p}\right>
=\mathbf{1}\left<\mathbf{p}\right>
=\mathbf{p}
\end{align*}
We hereafter recognize  $\mathbf{H}$ as the production networks comprising the entire potential alternative technologies, or  the meta structure, while $\nabla \mathbf{H}$ representing the equilibrium production network in terms of shares of the inputs for all sectors (i.e., input--output coefficient matrix).


\subsubsection{Networking Clusters in Equilibrium}
\begin{figure}[t!]
\includegraphics[width=.235\textwidth]{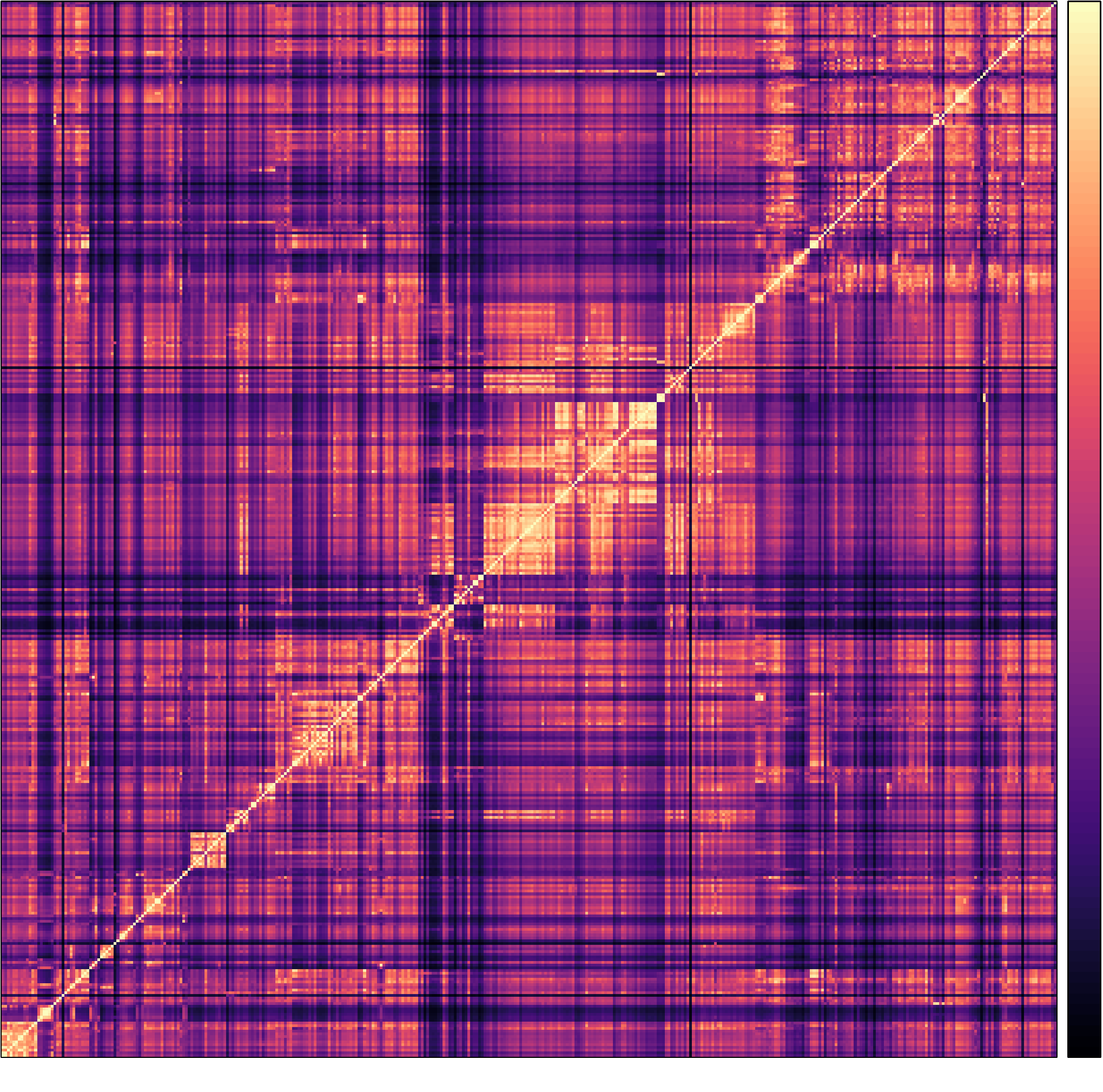}
\includegraphics[width=.235\textwidth]{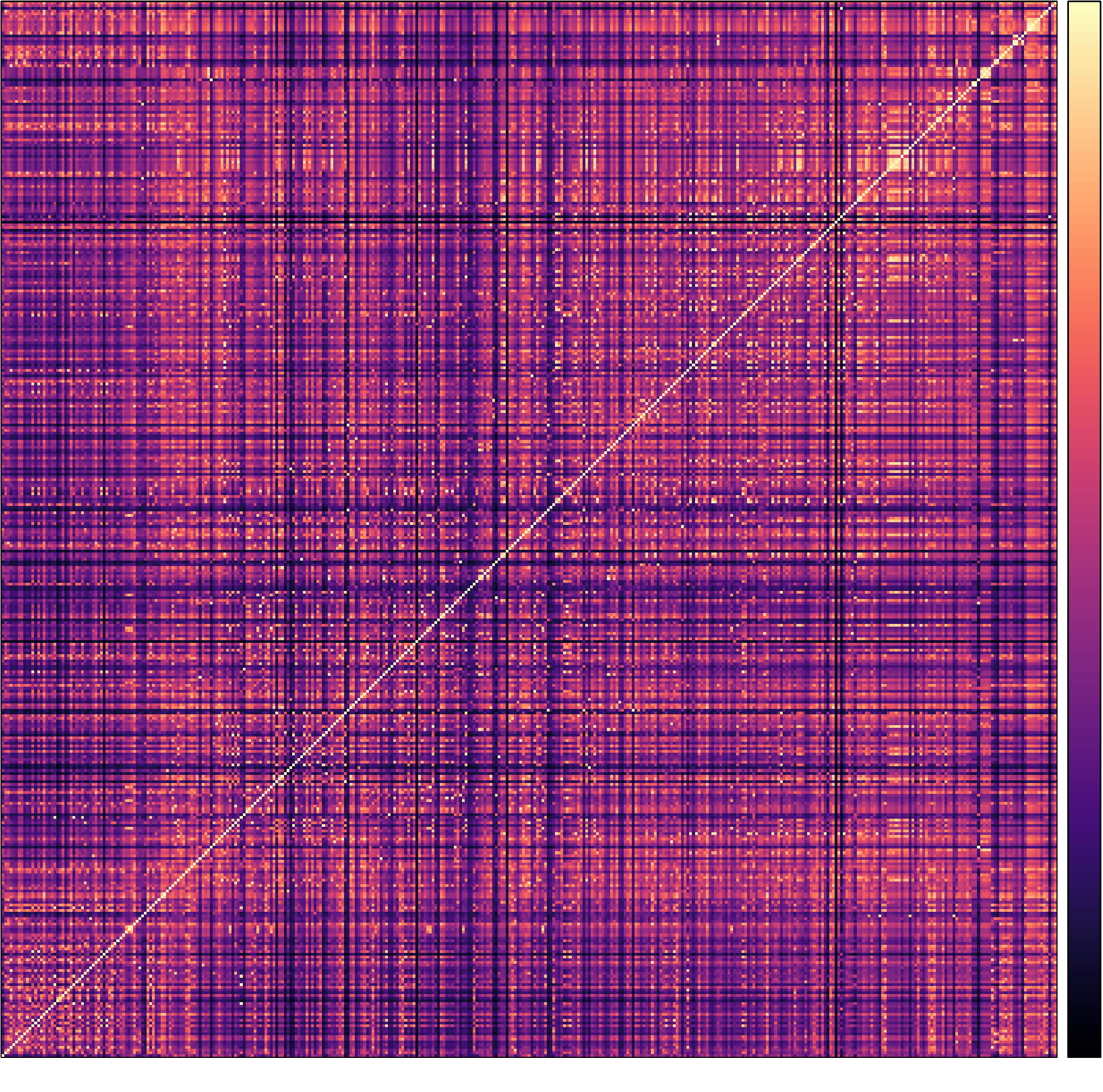}%
\caption{Scaled Euclidean distances $d_{jk}$ between net multipliers of industries $(\mu_j, \mu_k)$ in original order (left) and in stream order (right), for the current state.\label{heat}}
\end{figure}
In order to describe the production network in different equilibrium states, 
as well as to make comparisons between them, we perform cluster analysis.
Technological similarity of sectors in a production network has been studied by various methods pertaining to the measurement of distances between a pair of sectors \cite{jmn, pre2011, newman}.
In the present study, we measure cluster distance based on the correlation between the net \text{multipliers} of two sectors.
The net multiplier $\mu_j$ of sector $j$ measures the indirect requirements from all sectors needed to deliver a unit output to final demand from sector $j$.
Specifically, $\mu_j$ is an $n$-column vector of the following identity:
\begin{align*}
\left[ {\mu}_1, \cdots, \mu_n \right] = \left[ \mathbf{I} - \mathbf{S} \right]^{-1} -\mathbf{I} 
= \mathbf{S} + \mathbf{S}^2 + \mathbf{S}^3 + \cdots
\end{align*}
where $\mathbf{S}$ represents the input--output coefficient matrix of a certain state.
In other words, $\mu_j$ portrays sector $j$'s characteristic pattern of unit output propagation.

In Fig.~\ref{heat}, we display the heatmap of the correlations between all possible pairs of the net multipliers $(\mu_{j}, \mu_{k})$ for the current state.
Note that we convert Pearson correlations into the following scaled Euclidean distance: 
\begin{align*}
d_{jk} = \sqrt{1 - \text{Corr}\left(\mu_{j}, \mu_{k} \right)}
\end{align*}
Thus, a perfect correlation has zero distance, whereas a zero correlation has a distance of $1$ and a perfect negative correlation has a distance of $\sqrt{2}$.
The greatest distance shown in Fig.~\ref{heat} is indicated by the darkest color.
Notice that clustering is observable in the original order of the sector classification.
This is supposedly because the sectoral classification of an input--output table is based fundamentally on Colin Clark's three-sector theory, and the sectors are disaggregated while placed in the neighborhood.
On the other hand, the clusters are homogenized  more or less in the case of stream order as far as Fig.~\ref{heat} is concerned.

\subsubsection{Networking Clusters Transformation}
\begin{figure}[t!]
\includegraphics[width=.46\textwidth]{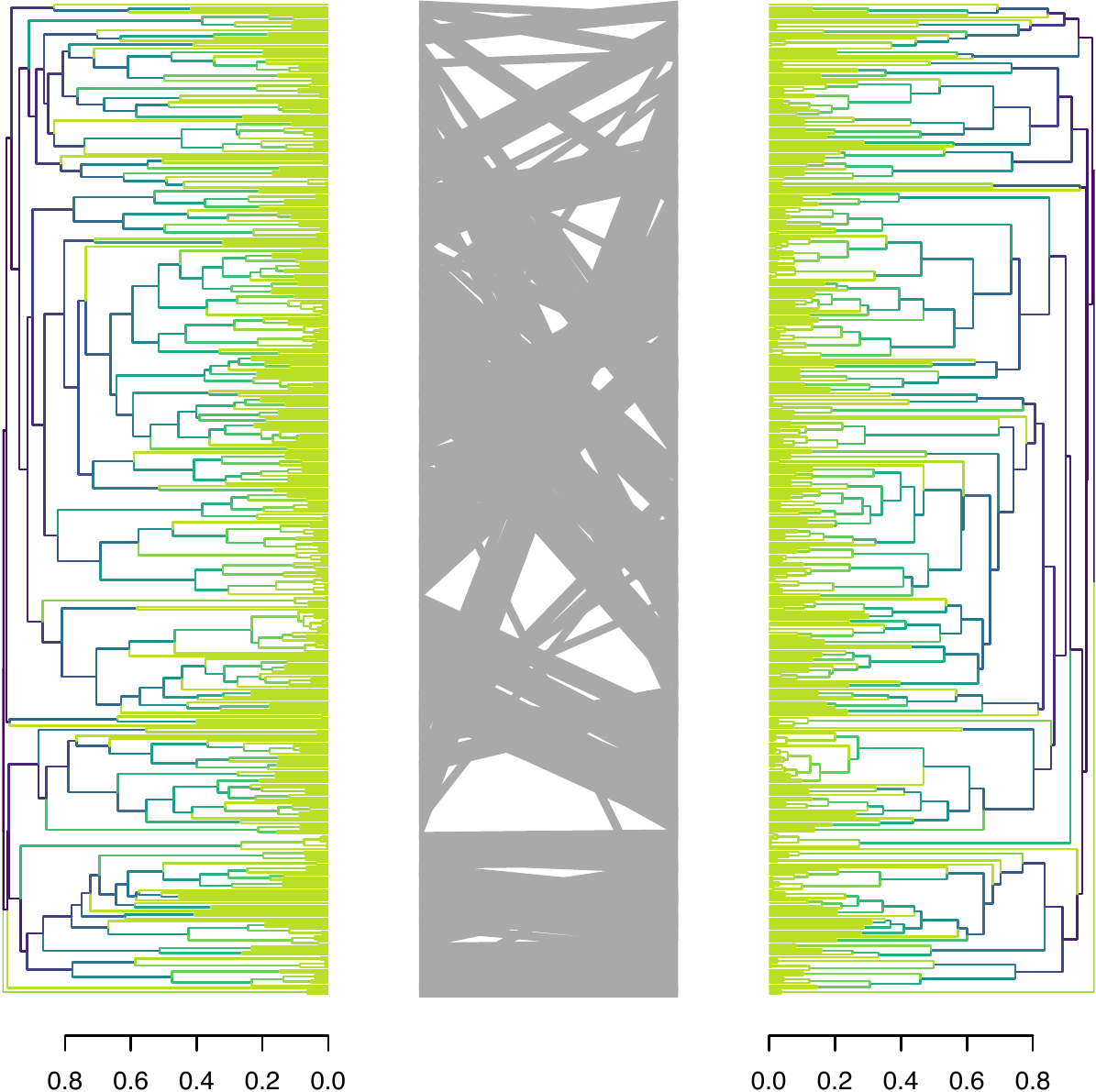}%
\caption{
Hierarchical clustering dendrograms by the scaled Euclidean distances $d_{jk}$ between the net multipliers of sectors for the reference (left) and the current (right) states.
Each leaf of a tree corresponds to a sector.
Network transformation 
regarding 
$d_{jk}$
can be monitored by the difference in the order of sectors configured by the dendrogram.
The same leaves (sectors) of the two trees are connected by a gray line.
\label{ddend1}}
\end{figure}

\begin{figure}[t!]
\includegraphics[width=.48\textwidth]
{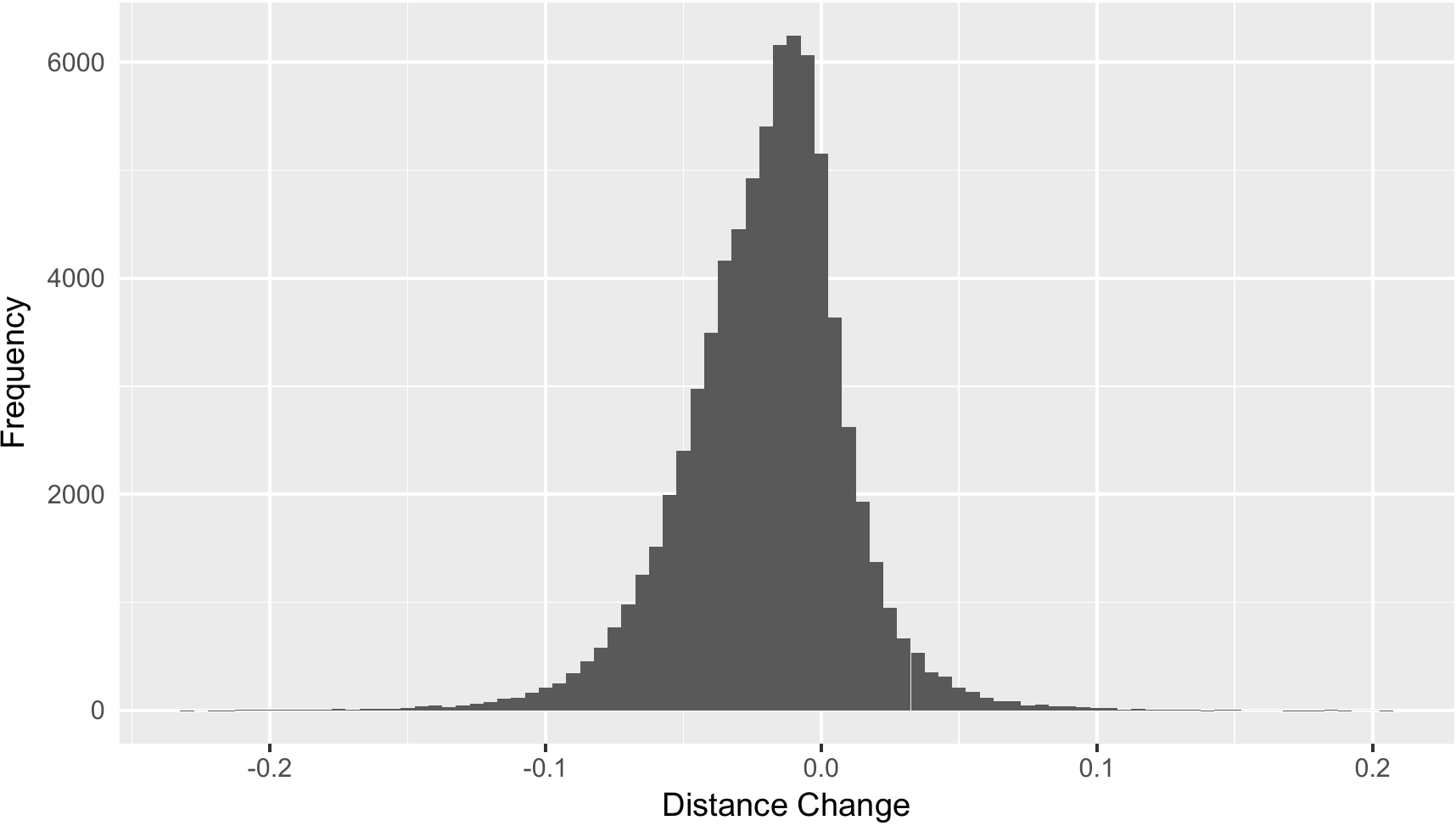}
\caption{
Distribution of changes in scaled Euclidean distances between the net multipliers of $\mathbf{A}$ and $\mathbf{B}$.
\label{hist21}}
\end{figure}

We further examine the changes in the networking clusters in different  states.
However, we found that the changes  in the multiplier correlations between different equilibrium states were visually undetectable by way of a heatmap.
Hence, we examine the transformation of the networking clusters  by way of the changes  in the correlation distances between all possible pairs of the multipliers.
In Fig.~\ref{ddend1}, we show the results of hierarchical clustering using dendrograms for the reference and current distance metrics of the correlations between the net multipliers.
In Fig.~\ref{hist21}, we display a histogram of the differences of distance between the two states.
Overall, the distances have contracted in the current state relative to the reference state.

Given the meta structure $\mathbf{H}$, we may project  the equilibrium state, given an exogenous productivity shock $\mathbf{z}=\left( z_1, \cdots, z_n \right)$.
The projected equilibrium price $\boldsymbol{\pi} = \left(\pi_1, \cdots, \pi_n \right)$ is the solution to the following system of equations:
\begin{align*}
\boldsymbol{\pi} = \mathbf{H}\left( \boldsymbol{\pi}, p_0 \right) \left< \boldsymbol{\theta} \right>^{-1} \left< \mathbf{z} \right>^{-1}
\end{align*}
The solution can be obtained by iteration, if the given productivity shock is enhancing, i.e., $z_j \geq 1$ for all $j$, since the domain will be contracting during this iteration process.
The iteration begins with the current equilibrium price; thus, the initial $(k=0)$ guess of the equilibrium price is $\mathbf{w}^{(0)}=\mathbf{p}$. 
The recurrence formula for $\mathbf{w}$ can be described as follows:
\begin{align}
\mathbf{w}^{(k+1)}=\mathbf{H}\left( \mathbf{w}^{(k)}, p_0 \right) \left< \boldsymbol{\theta} \right>^{-1} \left< \mathbf{z} \right>^{-1}
\label{contraction}
\end{align}
Specifically, the domain $(\mathbf{0}, \mathbf{p})$ will be contracted
\begin{align*}
\mathbf{p}
\geq
\mathbf{p}\left< \mathbf{z} \right>^{-1}
&=\mathbf{H}\left( \mathbf{p}, p_0 \right) \left< \boldsymbol{\theta} \right>^{-1} \left< \mathbf{z} \right>^{-1} 
\\
&\geq \mathbf{H}\left( \mathbf{0}, p_0 \right) \left< \boldsymbol{\theta} \right>^{-1} \left< \mathbf{z} \right>^{-1} > \mathbf{0}
\end{align*}
Hence, (\ref{contraction}) is a contraction mapping that converges monotonically to the equilibrium such that $\mathbf{p} \geq \mathbf{w}^{(k)} \rightarrow \boldsymbol{\pi} > \mathbf{0}$.
Note that we only used the monotonicity of $\mathbf{H}$ to show the convergence of the process (\ref{contraction}) for the enhancing case (i.e., $\mathbf{z}\geq \mathbf{1}$).
For a non-enhancing case, we need a guarantee that an equilibrium exists, which is provided if $\mathbf{H}$ is a concave mapping \cite{kras}; we may then take any initial guess to arrive at the equilibrium via (\ref{contraction}).

The social benefit of innovation (in terms of enhanced productivity $\mathbf{z}>\mathbf{1}$) relative to  the current state must be the reduced projected price of commodities $\boldsymbol{\pi} < \mathbf{p}$.
In this study, we evaluate such social benefit by the increase in earnable final demand for the same total amount of the primary input.
In other words, we evaluate innovation in terms of how much more net output can be provided from the same total amount of primary input.
For later convenience, we introduce the projected input--output coefficients ex post of $\mathbf{z}$ over  current state as follows:
\begin{align*}
\mathbf{m}_0 &= p_0 \nabla_0 \mathbf{H}\left( \boldsymbol{\pi}, p_0 \right) \left<\boldsymbol{\pi} \right>^{-1} \left< \boldsymbol{\theta}\right>^{-1} \left< \mathbf{z} \right>^{-1} \\
\mathbf{M}&= \left< \boldsymbol{\pi} \right> \nabla \mathbf{H}\left( \boldsymbol{\pi}, p_0 \right) \left<\boldsymbol{\pi} \right>^{-1}\left<\boldsymbol{\theta} \right>^{-1}\left< \mathbf{z} \right>^{-1} 
\end{align*}
Then, our innovation assessment problem can be specified as the following:
\begin{align*}
\max_{\delta}~\mathbf{1} \mathbf{f} \delta ~~\text{s.t.}~~ \mathbf{m}_0 \left[ \mathbf{I} - \mathbf{M} \right]^{-1} \left<\boldsymbol{\pi}\right>  \mathbf{f} \delta \leq \mathbf{b}_0 \left[ \mathbf{I} - \mathbf{B} \right]^{-1} \mathbf{f}
\end{align*}
where $\mathbf{f}$ is the current-state net output (or final demand) observed in the form of a  column vector, while $\delta$ is the scalar to be maximized.
Note that the right-hand side of the constraint is the total of primary inputs of the current state, and the objective term is the total earnable final demand at the projected state whose commodity-wise proportion is fixed at the current state.
The solution of the problem can be obtained by the following calculation:
\begin{align*}
\delta^* = \frac{\mathbf{b}_0 \left[ \mathbf{I} - \mathbf{B} \right]^{-1} \mathbf{f}}{\mathbf{m}_0 \left[ \mathbf{I} - \mathbf{M} \right]^{-1} \left<\boldsymbol{\pi}\right> \mathbf{f}}
\end{align*}
Further, we may examine the distribution of the primary inputs in the current and projected states, as follows:
\begin{align*}
\Delta \mathbf{v}
=
\mathbf{b}_0 \left[ \mathbf{I} - \mathbf{B} \right]^{-1} \left< \mathbf{f} \right>
-
\mathbf{m}_0 \left[ \mathbf{I} - \mathbf{M} \right]^{-1} \left<\boldsymbol{\pi}\right> \left< \mathbf{f} \delta^* \right>
\end{align*}
Here, $\Delta \mathbf{v}$ denotes the redistribution of the primary inputs, whose entries will sum to zero.
Moreover, we may calculate the economic welfare gain provided by $\mathbf{z}$, as the gain in the final demand, as follows:
\begin{align*}
\Delta f = \mathbf{1} \Delta\mathbf{f}=\mathbf{1}\mathbf{f}\left(\delta^* - 1\right)
\end{align*}
Note that $\mathbf{1f}$ is the aggregated demand, which equals the GDP of the current state economy.

\subsubsection{Structural Propagation of RMC110}
\begin{figure}[t!]
\includegraphics[width=.46\textwidth]{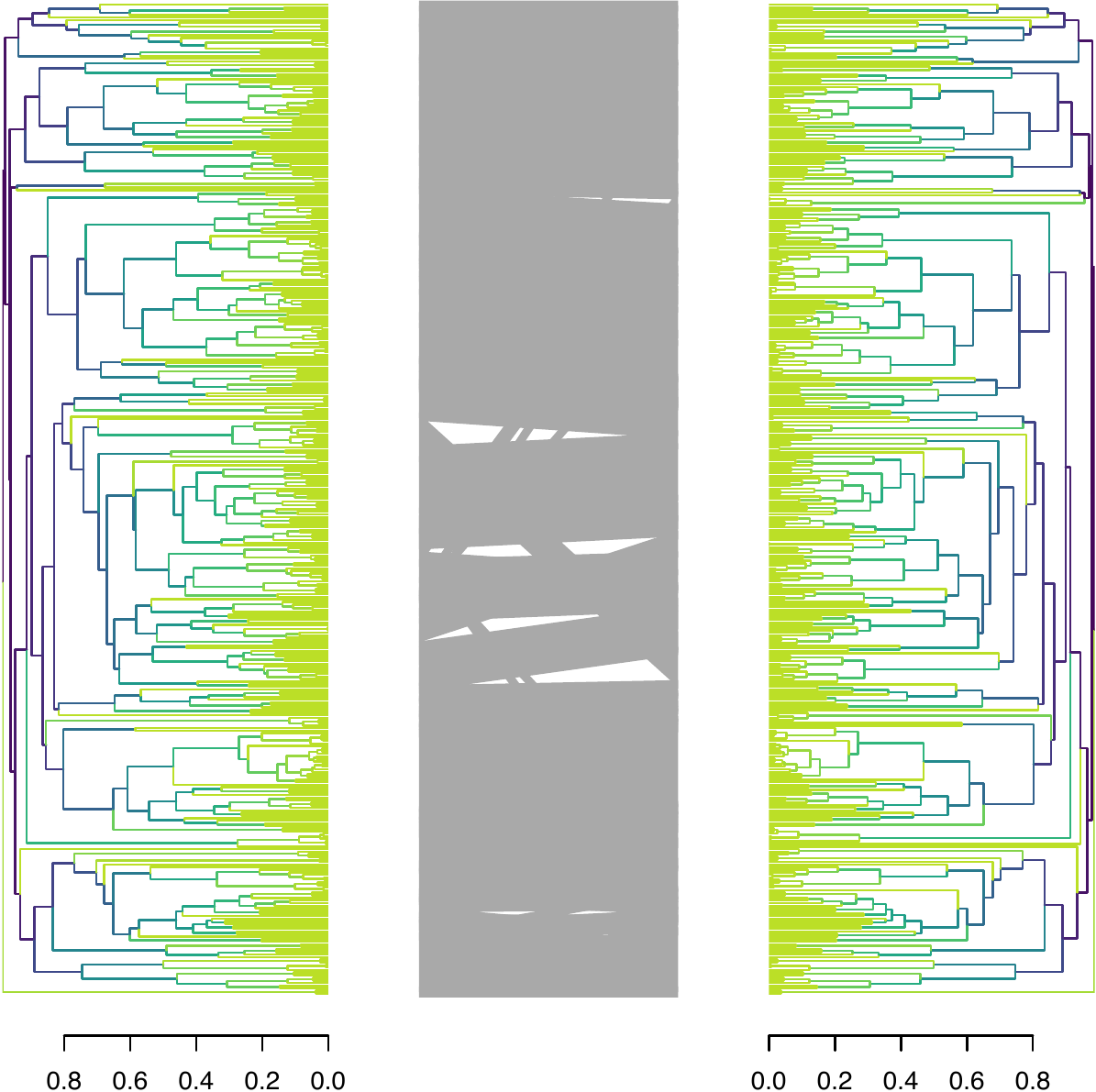}%
\caption{
Hierarchical clustering dendrograms by 
$d_{jk}$ between the net multiplier of sectors for the current state (left) and the projected state given by RMC110 (right).
The left tree is identical with the right tree of Fig. \ref{ddend1}
\label{ddend2}}
\end{figure}

\begin{figure}[th!]
\includegraphics[width=.48\textwidth]
{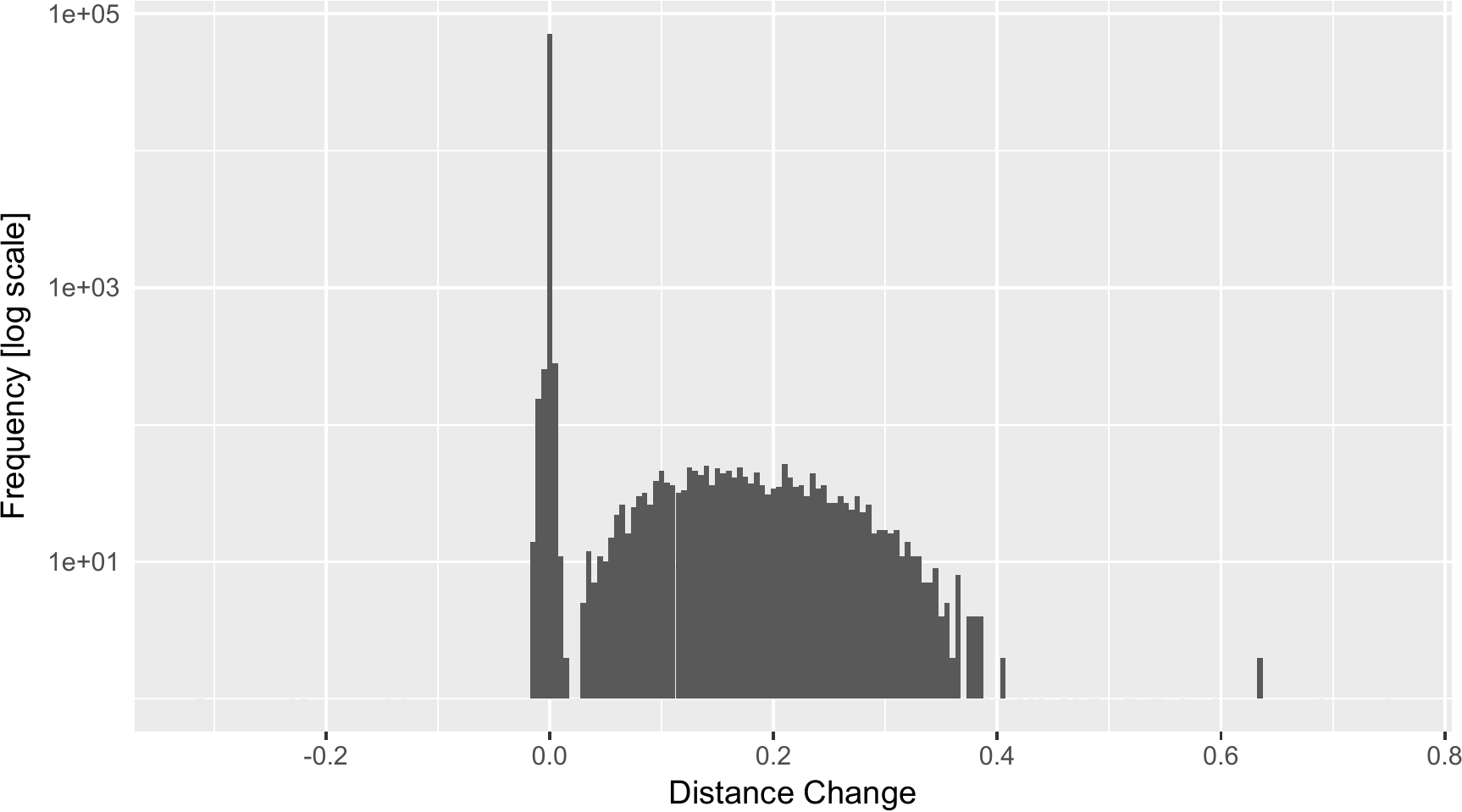}
\caption{
Distribution of changes in scaled Euclidean distances between the net multipliers of $\mathbf{B}$ and $\mathbf{M}$.
\label{hist32}}
\end{figure}

Here, we examine the structural propagation effect of an RMC110 innovation (where ``RMC'' refers to the ``Ready mixed concrete''  sector and ``110'' indicates a 10\% productivity increase)
injected into the current state economy.
RMC110 is specifically the following vector:
\begin{align}
&\mathbf{z}=\left( 1, \cdots, 1, z_{\text{RMC}}, 1, \cdots, 1\right), & z_{\text{RMC}} &= 1.10
\label{trigger}
\end{align}
Note that RMC appears 145th in the original order and 244th in the (upstream-first) stream order.
The empirically observed T{\"o}rnqvist index value is approximately $1.01$ for the interval between the reference and current states.
By recursive means, $\boldsymbol{\pi}$ is obtained via (\ref{contraction}) under  
(\ref{trigger}), and accordingly, the ex post equilibrium structure ($\mathbf{m}_0, \mathbf{M}$) is obtained as well.
We calculate the scaled Euclidean distances between the net multipliers of $\mathbf{M}$ and compare them with those of $\mathbf{B}$ in Fig.~\ref{hist32}.
Since RMC110 is a very small injected productivity shock \footnote{Japan's GDP of the \text{current} state was $467,159$ BJPY, while the output of the RMC sector was $1,134$ BJPY. 
Thus, 10\% of the current RMC output amounts to 0.0243\% of the current GDP. }, the scaled Euclidean distances of the output multipliers have changed just slightly towards sparsity.
In Fig.~\ref{ddend2}, we show the tanglegram comprising the results of hierarchical clustering for the current and projected distance metrics of the correlations between the multipliers. 
Note that the right-hand tree of Fig.~\ref{ddend1} is identical to the left-hand tree of Fig.~\ref{ddend2}.
The clusters have transformed  to a certain extent between the reference and current states, as well as slightly between the current state and the projected state given by RMC110.

\begin{figure}[t!]
\includegraphics[width=0.48\textwidth]{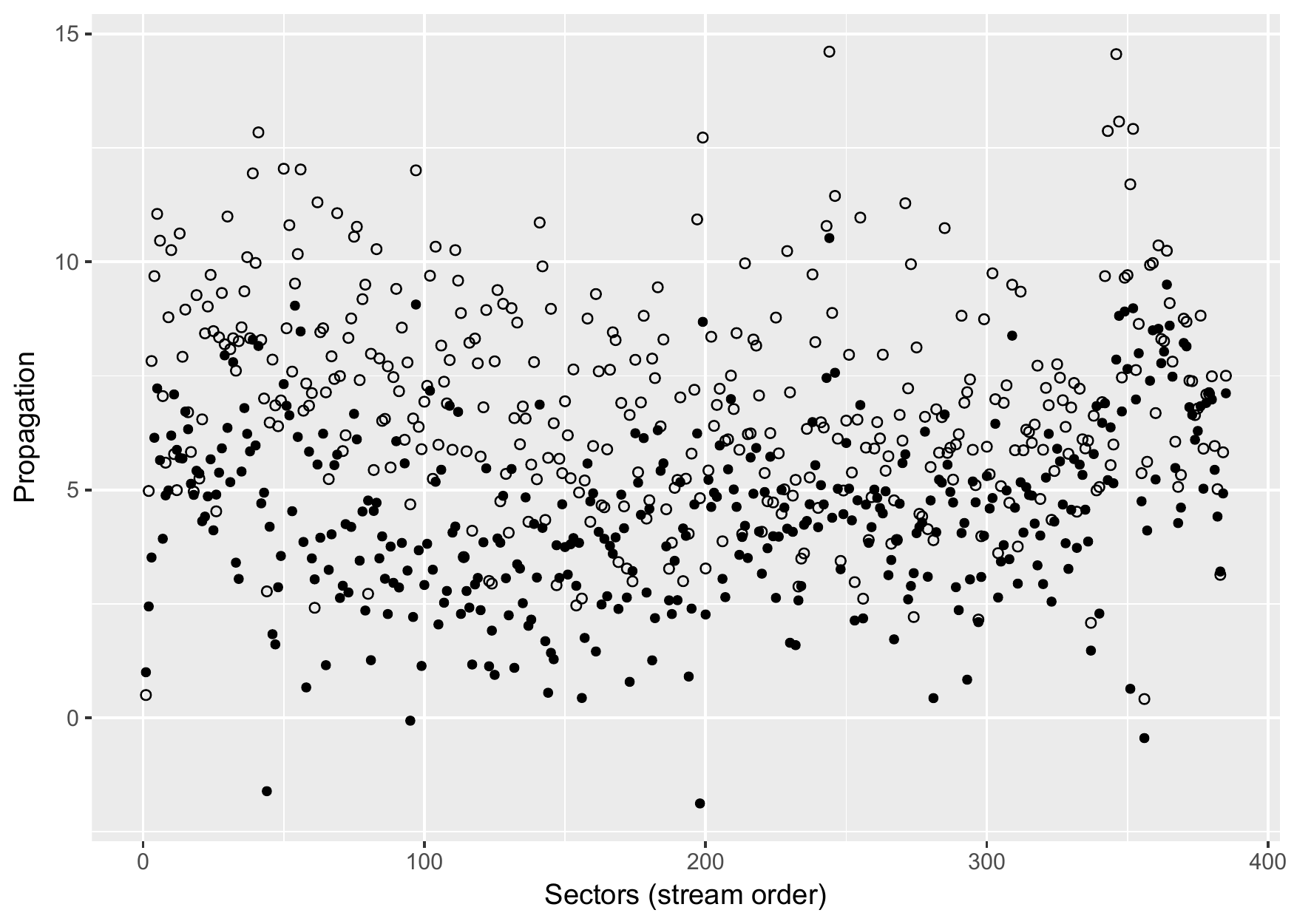}
\caption{
Each dot represents the sectoral propagation of RMC110 in terms of log-absolute difference between current and projected primary inputs i.e., $\ln \|\Delta v_j \|$.
Open dots correspond to structural propagation under Cascaded CES whereas filled dots correspond to non-structural propagation under Leontief networks.
\label{larmc110}}
\end{figure}

Figure \ref{larmc110} displays the log-absolute difference between current and projected primary inputs $\ln \| \Delta v_j \|$ in stream order, under Cascaded CES (open dots) and Leontief (filled dots) networks.
Note that Leontief network eliminates any sort of technology substitution (i.e., assuming $\sigma_{ij}=0$ for all $i$ and $j$). 
Structural propagation under Cascaded CES network clearly outperforms non-structural propagation under Leontief meta structure.

Table \ref{tabx} summarizes the welfare calculation with regard to structural propagation of RMC110 applied to the current state. 
Here, $y_\text{RMC}$ and $v_{\text{RMC}}$ indicate gross output and value added, respectively, of the RMC sector.   
The breakdown of $v_{\text{RMC}}$ indicates that primary inputs are to be redistributed to sectors such as Engine, Ready mixed concrete, and Wholesale trade, from sectors such as Public construction of roads, Miscellaneous civil engineering, and Civil engineering and construction services, in order to gain an economic welfare equivalent to $+334,827$ MJPY, given by RMC110. 
Further, note that a crude evaluation of RMC110 is to note first that 10\% of the current state output of RMC amounts to $+113,414$ MJPY.
This is the amount that the RMC sector is able to gain initially from RMC110.
In contrast, the propagation effect that pertains to the total amount of welfare gain, even without technological substitution (Leontief), can be almost two-fold larger ($+206,452$), and would be three-fold larger ($+334,827$) if we were to consider the full propagation effect, inclusive of the potential structural change.

\begin{table}[t!]
\centering
\small
\caption{\label{tabx}
Evaluation of structural propagation of RMC110 (unit: million Japanese yen).
$y_{\text{RMC}}$ is the gross output of the RMC sector, which is the sum of 
final demand and the intermediate inputs total. $v_{\text{RMC}}$ is the value added 
of the RMC sector.
Overall, non-structural propagation (Leontief) gains twice the 10\% amount of the current gross output of RMC sector (i.e., $113,414$ MJPY), whereas structural propagation (with Cascaded CES) gains thrice this amount.
}
\begin{tabularx}{87mm}{LRRR}
\hline\noalign{\smallskip}
\textrm{unit: [MJPY]}&
\textrm{Current}&
\textrm{{Projected (Leontief)}}&
\textrm{{Projected (CCES)}
} 
\\
\hline\noalign{\smallskip}
$y_{\text{RMC}}$ &	$1,134,144$ & $1,033,056$ & $6,393,959$ \\
$v_{\text{RMC}}$ &	$416,835$ & $379,682$ & $2,350,857$ \\
$\Delta f$ & $\pm 0$ & $+206,452$ & $+334,827$  \\ \hline
\end{tabularx}
\end{table}

\section{Concluding Remarks \label{sec4}}
Unlike utility, production is a step-by-step practice.
Accordingly, we took into account the configuration of the processes that underlie a production activity.
We broke down the configuration of production into activities that comprised binary and nested processes.
For empirical purposes, we applied a CES function for each nest process, resulting in a Cascaded CES function for modeling industrial production. 
Moreover, we used the sequence of inputs obtained by the triangulated input--output incidence matrix for modeling the intra-sectoral sequence of processes, as we observed a stylized hierarchy among the intermediate processes spanning the empirical input--output transactions.

By using linked input--output tables as the two-point data source, the elasticity parameters of Cascaded CES functions were resolved synchronously though the calibration of productivity.
At the same time, substantial negative nest elasticity parameters and productivity growths were observed. 
These amounts being negative may be attributed to bias in price measurement in the presence of qualitative changes.
Still, these parameters caused no problems, as far as the general equilibrium analysis was concerned.
The credibility of the calibrated system is validated by the complete replication of the two observed states portrayed in the linked input--output tables.
Naturally, the analysis becomes more decisive  when we advance the study to work more on  capital and growth, quality considerations, and international trade, all of which remain for further investigation.

\appendix
\section{Two-stage Cascaded CES calibration \label{app-b}}
We start with a following two-stage Cascaded CES function, where $W_1 = {w}_0$.
\begin{align*}
c &= t^{-1}W_3 
\\
W_3 &= \left( \lambda_2 {w}_2^{1-\sigma_2} + \Lambda_2 W_{2}^{1-\sigma_2} \right)^{\frac{1}{1-\sigma_2}} \\
W_2 &= \left( \lambda_1 {w}_1^{1-\sigma_1} + \Lambda_1 W_{1}^{1-\sigma_1} \right)^{\frac{1}{1-\sigma_1}} 
\end{align*}
This is the two-stage ($n=2$) version of the function defined in (\ref{vncesucf}) and (\ref{cn}).
As regards (\ref{lamnk}), the share parameters are equalt to 
the cost shares at the reference state:
\begin{align*}
\lambda_{2} = a_2
,~~~~~~
\lambda_{1} = \frac{a_1}{1-a_2}
\end{align*}

\begin{figure}[t!]
\centering
\includegraphics[width=0.35\textwidth]{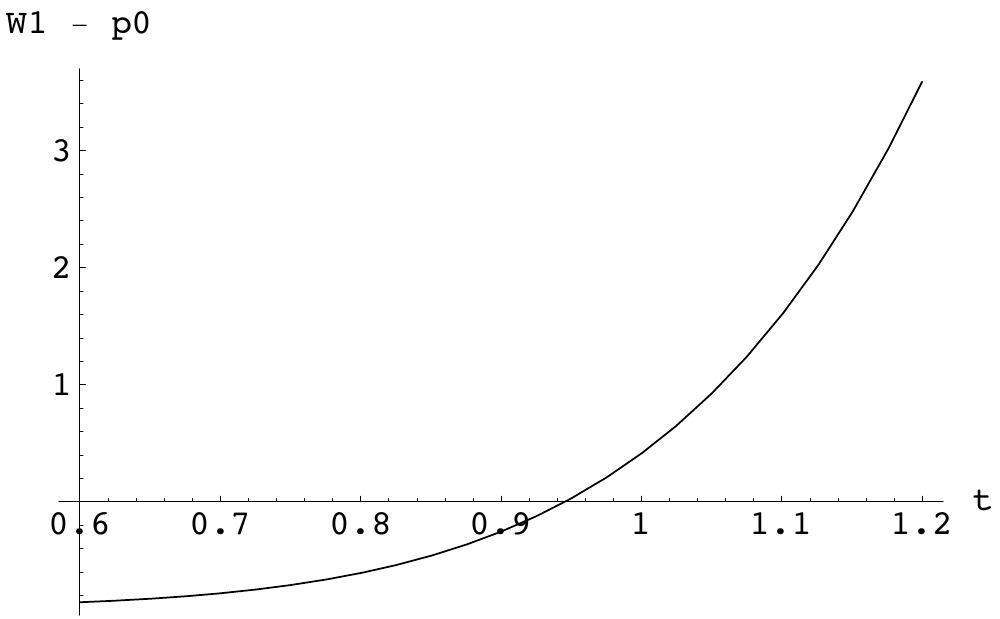}%
\caption{
Equation (\ref{calibr}) is solved at $t=0.946 \equiv \theta$.
\label{two}}
\end{figure}

The following exposition shows the procedure for parameter calibration via backward induction:
\begin{align*}
W_3 &= {t} {p} =W_3 \left( t; \chi_3 \right)  
\\
\sigma_2 
&
=\frac{\ln \frac{b_2}{a_2}+\ln \frac{W_3}{p_2}}{\ln \frac{W_3}{p_2}}
=\sigma_{2}\left( {t}; \chi_3, \chi_2 \right)
\\
W_2
&=\left(\frac{W_{3}^{1-\sigma_{2}} - {a}_{2} {p}_{2}^{1-\sigma_{2}}}{1-{a}_{2}} \right)^{\frac{1}{1-\sigma_{2}}}
=W_{2}\left( {t}; \chi_3, \chi_2 \right) 
\\
\sigma_1
&
= \frac{\ln \frac{b_1}{a_1} + \sigma_2 \ln \frac{W_2}{W_3} +\ln \frac{W_3}{{p}_1} }{\ln \frac{W_2}{p_1}}
=\sigma_{1}\left( {t}; \chi_3, \chi_2, \chi_1 \right) 
\\
W_1 &= 
\left(\frac{W_{2}^{1-\sigma_{1}} - {a}_{1} {p}_{1}^{1-\sigma_{1}}}{1 - {a}_1} \right)^{\frac{1}{1-\sigma_{1}}}
=W_1\left( {t}; \chi_3, \chi_2, \chi_1 \right)
\end{align*}
Hence, we can calibrate $t$ for the final condition, which is the two-stage version of (\ref{w1p0}):
\begin{align}
W_1\left( {t}; \chi_3, \chi_2, \chi_1 \right) = {p}_0
\label{calibr}
\end{align}
For demonstration purposes, we calibrate $t$ on the data given in Table \ref{tab:exam}.
We use the data, i.e.,  ${\chi}_i= \left\{ a_i, b_i, {p}_i \right\}$ and $p_0$ for calibrating $t$.
Note that $a_0$ and $b_0$ are not used in the calculation, because the shares are degenerate, i.e., $\sum_{i}s_i =1$.
Figure \ref{two} illustrates how (\ref{calibr}) is solved for $t$; the solution $t=0.946 \equiv \theta$ is listed in Table \ref{tab:exam}.  
We also include TFPg calculable from the same data, using (\ref{tl}).
The elasticity parameters $\sigma_1$ and $\sigma_2$ are resolved at the calibrated productivity $\theta$.
\begin{table}[t!]
\centering
\small
\caption{Sample data (shaded values correspond to $\chi_3$, $\chi_2$, $\chi_1$ and $p_0$) and the calibrated parameters. \label{tab:exam}}
 \begin{tabularx}{86mm}{lCCCCcc}
\hline\noalign{\smallskip}
 &	$a$ &	$b$ &	${p}$ & $\sigma_i$ &	$\theta$&T{\"o}rnqvist\\\hline\noalign{\smallskip}
output&&&\cellcolor{gray!20}{0.8}&&$0.946$ & $0.947$	\\	
input 2&\cellcolor{gray!20}{0.3}&\cellcolor{gray!20}{0.2}&\cellcolor{gray!20}{1.2}&1.88	&&	 \\	
input 1&\cellcolor{gray!20}{0.5}&\cellcolor{gray!20}{0.7}&\cellcolor{gray!20}{0.6}&	3.54&	& \\	
input 0&{0.2}&{0.1}&\cellcolor{gray!20}{0.9}&&&\\	\hline\noalign{\smallskip}
\end{tabularx}
\end{table}

\section*{References}
\raggedright
\bibliography{bibNN}

\end{document}